\documentclass[aps,pra,twocolumn,superscriptaddress,showpacs]{revtex4}

\pdfoutput=1

\usepackage{amsmath,graphicx,amssymb,braket,xcolor,subfigure,upgreek,bbold}
\usepackage[colorlinks, linkcolor=blue, citecolor=blue, urlcolor=blue, breaklinks=true]{hyperref}
\usepackage{verbatim}
\usepackage{float}
\usepackage{esint}

\newcommand{\be}{\begin{equation}}
\newcommand{\ee}{\end{equation}}
\newcommand{\ba}{\begin{aligned}}
\newcommand{\ea}{\end{aligned}}

\begin{document}

\title{Entanglement growth and correlation spreading with variable-range interactions in spin and fermionic tunnelling models}

\author{Anton S. Buyskikh}
\affiliation{Department of Physics and SUPA, University of Strathclyde, Glasgow G4 0NG, UK}
\author{Maurizio Fagotti}
\affiliation{D\'epartement de Physique, \'Ecole Normale Sup\'erieure / PSL Research 
University, CNRS, 24 rue Lhomond, 75005 Paris, France}
\author{Johannes Schachenmayer}
\affiliation{JILA, NIST, Department of Physics, University of Colorado, 440 UCB, Boulder, CO 80309, USA}
\author{Fabian Essler}
\affiliation{The Rudolf Peierls Centre for Theoretical Physics,
Oxford University, Oxford, OX1 3NP, UK}
\author{Andrew J. Daley}
\affiliation{Department of Physics and SUPA, University of Strathclyde, Glasgow G4 0NG, UK}

\date{March 17, 2016}


\begin{abstract}
We investigate the dynamics following a global parameter quench for two 1D models with variable-range power-law interactions: a long-range transverse Ising model, which has recently been realised in chains of trapped ions, and a long-range lattice model for spinless fermions with long-range tunnelling. For the transverse Ising model, the spreading of correlations and growth of entanglement are computed using numerical matrix product state techniques, and are compared with exact solutions for the fermionic tunnelling model. We identify transitions between regimes with and without an apparent linear light cone for correlations, which correspond closely between the two models. For long-range interactions (in terms of separation distance $r$, decaying slower than $1/r$), we find that despite the lack of a light-cone, correlations grow slowly as a power law at short times, and that -- depending on the structure of the initial state -- the growth of entanglement can also be sublinear. These results are understood through analytical calculations, and should be measurable in experiments with trapped ions. 
\end{abstract}

\maketitle

\section{Introduction}

In recent years, advances in experiments with trapped ions \cite{Schneider2012,Islam2011,Kim2011,Lanyon2011,Blatt2012}, ultracold polar molecules \cite{Yan2013,Micheli2006}, and Rydberg atoms \cite{Schausz2012,Weimer2010,Pohl2010} have allowed for the experimental realisation of highly-controllable spin models with interactions that decay as a power law. Ions in Paul and Penning traps, in particular, offer an opportunity to design spin models with variable range interactions by mediating interactions between internal spin states via the collective motional modes of the ions \cite{Porras2004,Deng2005}. Parameters in these models can be controlled -- and the spin dynamics measured --- time-dependently, opening new opportunities to study quench dynamics in closed quantum systems \cite{Jurcevic2014,Richerme2014,Schachenmayer2013,Hauke2013,Gong2013,Britton2012}. 

This has opened new fundamental questions related to the propagation of correlations in such systems -- in particular, how to generalise the Lieb-Robinson bounds \cite{Lieb1972} for the spreading of correlations in systems with nearest-neighbour interactions to situations with long-range interactions. While a series of general results have been derived, allowing gradually tighter bounds \cite{Hastings2006,Eisert2013,Gong2014,Storch2015,Foss-Feig2015,Maghrebi2015,Schachenmayer2015,Rajabpour2014}, it is particularly informative to identify exactly solvable models that reproduce and explain the qualitative behaviour of the physical systems being studied. Up to now there are relatively few models with long-range interactions (limited mainly to longitudinal Ising models \cite{Hazzard2014} and tunnelling bosons \cite{Cevolani2015, Tezuka2014}) for which the transitions in behaviour are known. 

In this article, we study global quench dynamics in two models with long range interactions. The first is the transverse Ising model with long-range interactions \cite{Porras2004,Deng2005}, which was recently realised in a series of experiments \cite{Islam2011,Kim2011,Jurcevic2014,Richerme2014}, but for which no general analytical solution is known. The second model is an exactly solvable model that represents a generalisation of a Kitaev chain \cite{Kitaev2001} with long-range tunnelling. For moderate system sizes and timescales, we compute the dynamics of the long-range transverse Ising model numerically by using Matrix Product Operator (MPO) techniques \cite{Verstraete2008,Garcia-Ripoll2006,McCulloch2007,Pirvu2010,Crosswhite2008}.
For systems with long-range interactions, these allow a particularly convenient implementation of propagation schemes for matrix product states (MPS), the state representation that forms the basis for the time-dependent density matrix renormalisation group methods \cite{Schollwoeck2011, Schollwoeck2005, White2004, Daley2004}. We compare the qualitative behaviour of the two models in terms of spreading of correlations and growth of entanglement after a global quantum quench, beginning with a state that is initially uncorrelated. We also make use of a Holstein-Primakoff approximation to gain further insight into dynamics in the Ising model with long-range interactions, for short times and initial states with all of the spins aligned.

As was seen previously with local quenches in the transverse Ising model \cite{Hauke2013,Jurcevic2014}, we are able to classify the behaviour of correlation spreading in both of these models into different regimes as a function of the decay exponent $\alpha$ of power-law interactions (which decay as $1/r^\alpha$), where $r$ is the separation distance. For correlation spreading, the dynamics are divided into (i) a regime of short-range interactions where $\alpha>2$, (ii) a regime of intermediate and long-range interactions when $\alpha<2$, with certain features also changing at $\alpha=1$. While light-cone-like behaviour remains for intermediate an short-range interactions, in the case where $\alpha<1$, there is complete absence of a light cone for spreading of correlations.

Counterintuitively, though, in this regime the development of both correlations and entanglement can be suppressed. For the fermionic model, we find that the growth of correlations takes the form of a power law at long distances, leading to slow growth at short times. When the Hamiltonian and the initial state have similar symmetries, we also observe for both models that the growth of bipartite entanglement in spatial modes of the chain can be significantly suppressed at short times, in contrast to cases with shorter range interactions. We can understand this based on a change in the dispersion relation for long-range interactions, which diverges for modes with quasimomentum $k\rightarrow 0$. The initial state then affects the dynamics by determining to what extent new quasiparticles are produced with these momenta. 

The rest of this article is arranged as follows: In Sec.~\ref{sec:models}, we discuss the long-range transverse Ising and fermionic hopping models, giving details of the methods we use to solve these.  In Sec.~\ref{sec:correlations}, we then present the time-dependent correlation dynamics for each of the models after a quantum quench, discussing in each case the different regimes of short and long-range interactions. We then move on to discuss growth of entanglement in these models in Sec.~\ref{sec:entanglement}, before providing a summary and outlook in Sec.~\ref{sec:summary}. 

\section{Models with long-range interactions}
\label{sec:models}

In this section, we define the two systems with long-range interactions that we will analyse and compare. The first is the long-range transverse Ising model, and the second is an exactly solvable long-range hopping model for spinless fermions. 

\begin{figure}[tb]
\begin{centering}
\includegraphics[width=1\columnwidth]{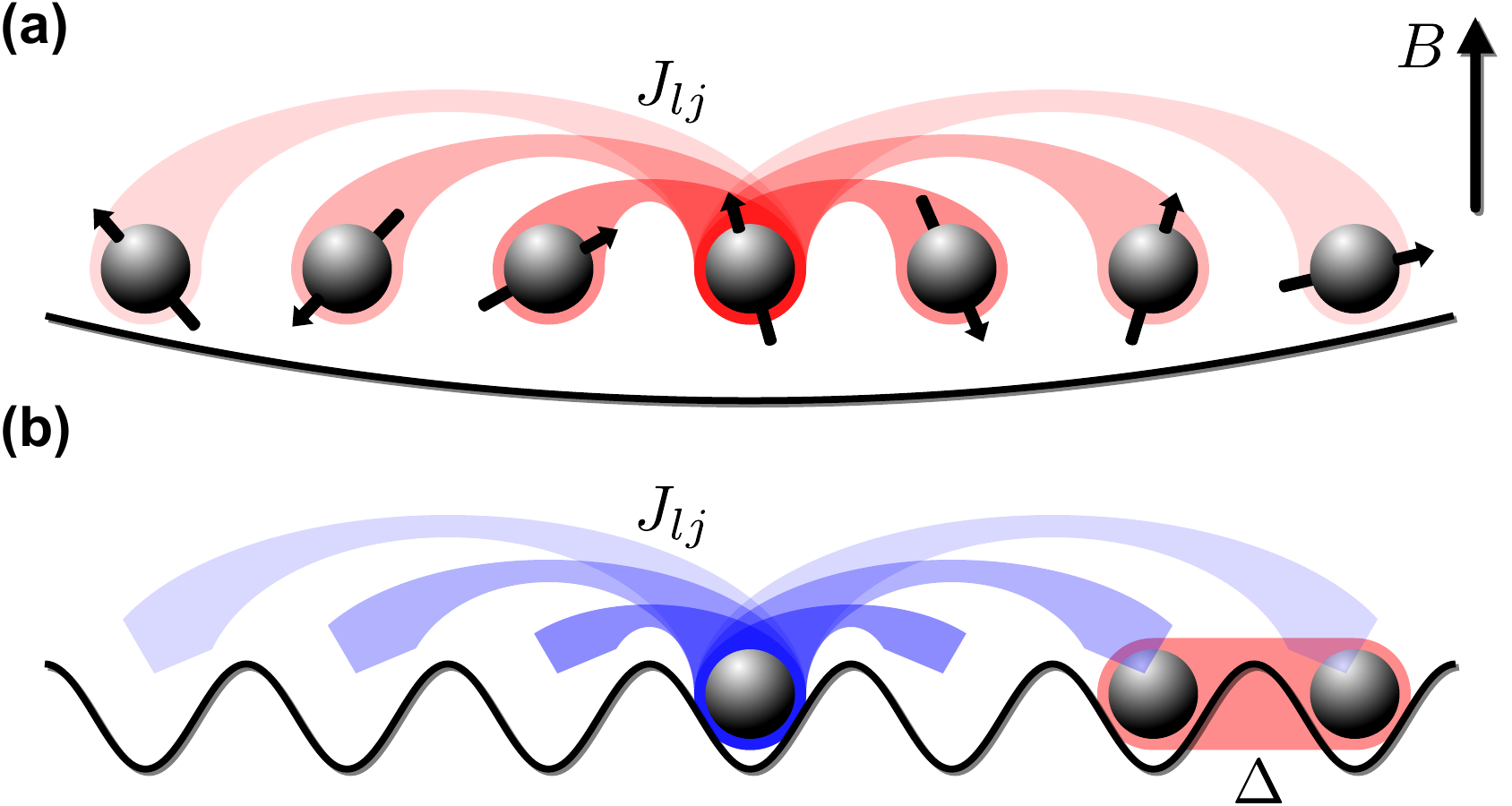}
\par\end{centering}
\caption{\label{fig:illustration} 
Illustration of the long-range models. (a) The long-range transverse Ising (LRTI) model, which has been realised experimentally in chains of trapped ions. The spins interact over a distance with amplitude $J_{lj}$, in the presence of a transverse field term $B$. (b) The long-range fermionic hopping (LRFH) model. Spinless fermions tunnel between distant sites in a 1D lattice with amplitude $J_{lj}$, and pairing on neighbouring sites is induced with strength $\Delta$.}
\end{figure}

\subsection{Long-range transverse Ising (LRTI) model}

The first model of interest is a 1D chain of $M$ spin-1/2 systems described by the long-range transverse Ising model ($\hbar\equiv 1$),
\begin{equation}
H_{\rm LRTI}=\sum_{l>j}^{M}J_{lj}\sigma_{l}^{x}\sigma_{j}^{x}+B\sum_{l=1}^{M}\sigma_{l}^{z},\label{eq:H_LRTI}
\end{equation}
where $J_{lj}$ is the spin-spin interaction matrix, $B$ is the transverse field, and $\sigma_l^{x,z}$ denote local Pauli matrices operating on each spin. It is possible to realise this Hamiltonian experimentally with 1D chains of trapped ions \cite{Porras2004}, as an effective model for the dynamics of long-lived internal states (denoted $\left|\uparrow\right\rangle _{l}$ and $\left|\downarrow\right\rangle _{l}$ for spin $l$) of the ions. The interactions are generated by cooperative spin-flips of the internal states, induced by external laser fields, and coupled via the motional modes of the ion chain. It is possible in experiments to achieve a close approximation to algebraic decay of correlations $J_{lj}=J\left|l-j\right|^{-\alpha}$, $J>0$, with open boundary conditions, and $0 \lesssim \alpha \lesssim 3$. By shifting the internal levels of individual ions an effective magnetic field $B_l$ can be realised,
and here we focus on a transverse uniform field.

Our primary original motivation for studying quench dynamics in this system is the easily realisable experimental sequence in which all of the spins are initially prepared in a single state (say, $\left|\downarrow\right\rangle _{l}$), and then dynamics are allowed to proceed under the Hamiltonian~\eqref{eq:H_LRTI}. This corresponds to a global quench from $B=\infty$ with the ground state $\left|\psi(t=0)\right\rangle =\otimes_{l}\left|\downarrow\right\rangle _{l}$ to some finite value of the field. 

This model does not have a known analytical solution, but we are able to compute the dynamics using exact diagonalisation for chain lengths up to $M\sim 20$, or for moderate times and longer chains up to $M\sim 100$ by applying time-dependent variational principle (TDVP) techniques with MPO representation of the Hamiltonian \cite{Verstraete2008,Garcia-Ripoll2006,McCulloch2007,Pirvu2010,Crosswhite2008,Haegeman2014}. For these calculations, the convergence in the MPS bond dimension $D$ and time step $\Delta t$ was tested to ensure accuracy of the calculations. Furthermore, we can qualitatively describe the short-time dynamics in the Holstein-Primakoff approximation (see Sec.~\ref{sec:HPapproximation}).

\subsection{Long-range fermionic hopping (LRFH) model}
\label{sec:LRFH_PBC_derivation}

The second system is a 1D lattice with spinless fermions that has the form of a generalised Kitaev chain model \cite{Kitaev2001} with long-range hopping \cite{vodola2015}
\begin{equation}
H_{\rm LRFH}=\sum_{l\neq j=1}^{M}\bar J_{lj}c_{l}^{\dagger}c_{j}+\Delta\sum_{l}c_{l}^{\dagger}c_{l+1}^{\dagger}+\mathrm{h.c.}.\label{eq:H_LRFH}
\end{equation}
Here, $c_{l}$ is a fermionic annihilation operator on site $l$, $\bar J_{lj}$ is the hopping matrix with long-range couplings analogous to those in the spin model, and $\Delta$ characterizes interactions between fermions in the pairing term. For analytical calculations in order to simplify expressions without the thermodynamic limit we will make use of periodic boundary conditions, and choose 
\begin{equation}
\bar J_{lj}=J\left|\frac{M}{\pi}\sin\left[\frac{\pi\left(l-j\right)}{M}\right]\right|^{-\alpha},
\end{equation}
where $J>0$. For numerical calculations we will follow the case of the spin model, and choose open boundary conditions with $\bar J_{lj}=J_{lj}$. We find that the behaviour in each case agrees well in the limit of large system sizes. Also, in analogy with the spin model, we will consider quantum quenches, typically starting from the ground state for large values of $\Delta$, and quenching to smaller values of $\Delta$. 

The quadratic Hamiltonian of Eq.~\eqref{eq:H_LRFH} can be diagonalized via Bogoliubov transformations, analogously to the recent results in Ref.~\cite{vodola2015}. In the case of periodic boundary conditions, the Hamiltonian in momentum space reads 
\begin{eqnarray*}
 H_{\rm LRFH}&=&\sum_{k=0}^{M/2-1}\left(\begin{array}{cc}
c_{k}^{\dagger} & c_{M-k}\end{array}\right)\times\\
 &  & \left(\begin{array}{cc}
\bar{\mathcal{J}}(k) & 2i\Delta\sin\left(\frac{2\pi k}{M}\right)\\
-2i\Delta\sin\left(\frac{2\pi k}{M}\right) & -\bar{\mathcal{J}}(k)
\end{array}\right)\left(\begin{array}{c}
c_{k}\\
c_{M-k}^{\dagger}
\end{array}\right),
\end{eqnarray*}
where
\begin{equation*}
c_k=\frac{1}{\sqrt{M}}\sum_{r=1}^M\mathrm{e}^{-i\frac{2\pi kr}{M}}c_r
\end{equation*}
and
\be 
\label{eq:Jk_PBC}
\bar{\mathcal{J}}(k)=2\sum_{d=1}^{M-1}J_{l,l+d}\cos\left(2\pi kd/M\right).
\ee
Then via the Bogoliubov transformations
\begin{eqnarray*}
\left(\begin{array}{c}
\alpha_{k}\\
\alpha_{M-k}^{\dagger}
\end{array}\right)=\left(\begin{array}{cc}
u & v\\
-v^{*} & u^{*}
\end{array}\right)\left(\begin{array}{c}
c_{k}\\
c_{M-k}^{\dagger}
\end{array}\right),
\end{eqnarray*}
where $u=\cos\left(\theta_k/2\right)$, $v=i\sin\left(\theta_k/2\right)$,
and the angle $\theta_k$ is chosen to satisfy
\begin{eqnarray*}
\mathrm{e}^{i\theta_k} & = & \frac{\bar{\mathcal{J}}(k) + 2i\Delta\sin\left(\frac{2\pi k}{M}\right)}{\sqrt{\bar{\mathcal{J}}^2(k) + 4\Delta^2 \sin^2\left(\frac{2\pi k}{M}\right)}},
\end{eqnarray*}
the Hamiltonian can be diagonalised, and is found to have eigenenergies $E(k)=\pm\sqrt{\bar{\mathcal{J}}^2(k)+4\Delta^2 \sin^2 \left( 2\pi k/M\right) } \equiv \pm \epsilon(k)$.

The pre-quenched Hamiltonian has another value $\Delta_{0}$, instead of $\Delta$, so the pre-quenched Bogoliubov particles and angle will be denoted $\alpha_k^0$ and $\theta_k^0$ respectively. The time evolution of the original fermionic operators will be
\begin{eqnarray*}
\left(\begin{array}{c}
c_{k}\left(t\right)\\
c_{M-k}^{\dagger}\left(t\right)
\end{array}\right)&=&\left(\begin{array}{cc}
\cos\left(\theta_k/2\right) & -i\sin\left(\theta_{k}/2\right)\\
-i\sin\left(\theta_{k}/2\right) & \cos\left(\theta_{k}/2\right)
\end{array}\right)\times \\
 &  & \left(\begin{array}{cc}
\mathrm{e}^{-i\epsilon(k)t} & 0\\
0 & \mathrm{e}^{i\epsilon(k)t}
\end{array}\right)\left(\begin{array}{c}
\alpha_{k}\\
\alpha_{M-k}^{\dagger}
\end{array}\right).
\end{eqnarray*}
From this, we can obtain the time evolution of the single particle density matrix after the quench, beginning with the ground state $\alpha_k^0\left| \phi_0 \right\rangle=0$:
\begin{eqnarray}
 \langle c_{l}^{\dagger}(t)c_{j}(t)\rangle &=&\frac{1}{2M}\sum_{k=0}^{M-1}\mathrm{e}^{-i\frac{2\pi k\left(l-j\right)}{M}}\left[1-\cos\theta_k \cos \delta\theta_k \right. \nonumber \\
 &  & \left.+\sin\theta_{k}\sin \delta\theta_k \cos \left( 2 \epsilon(k) t \right) \right],
 \label{eq:greens_function} 
\end{eqnarray}
where 
\be
\label{eq:deltatheta}
\delta\theta_{k}=\theta_k-\theta_k^0
\ee
is the difference of the Bogoliubov angles before and after the quench. In the following two sections we now look in detail at the spreading of correlations and the growth of entanglement, comparing the results of the LRTI model to the LRFH model.

\begin{figure}[tb]
\begin{centering}
\includegraphics[width=1\columnwidth]{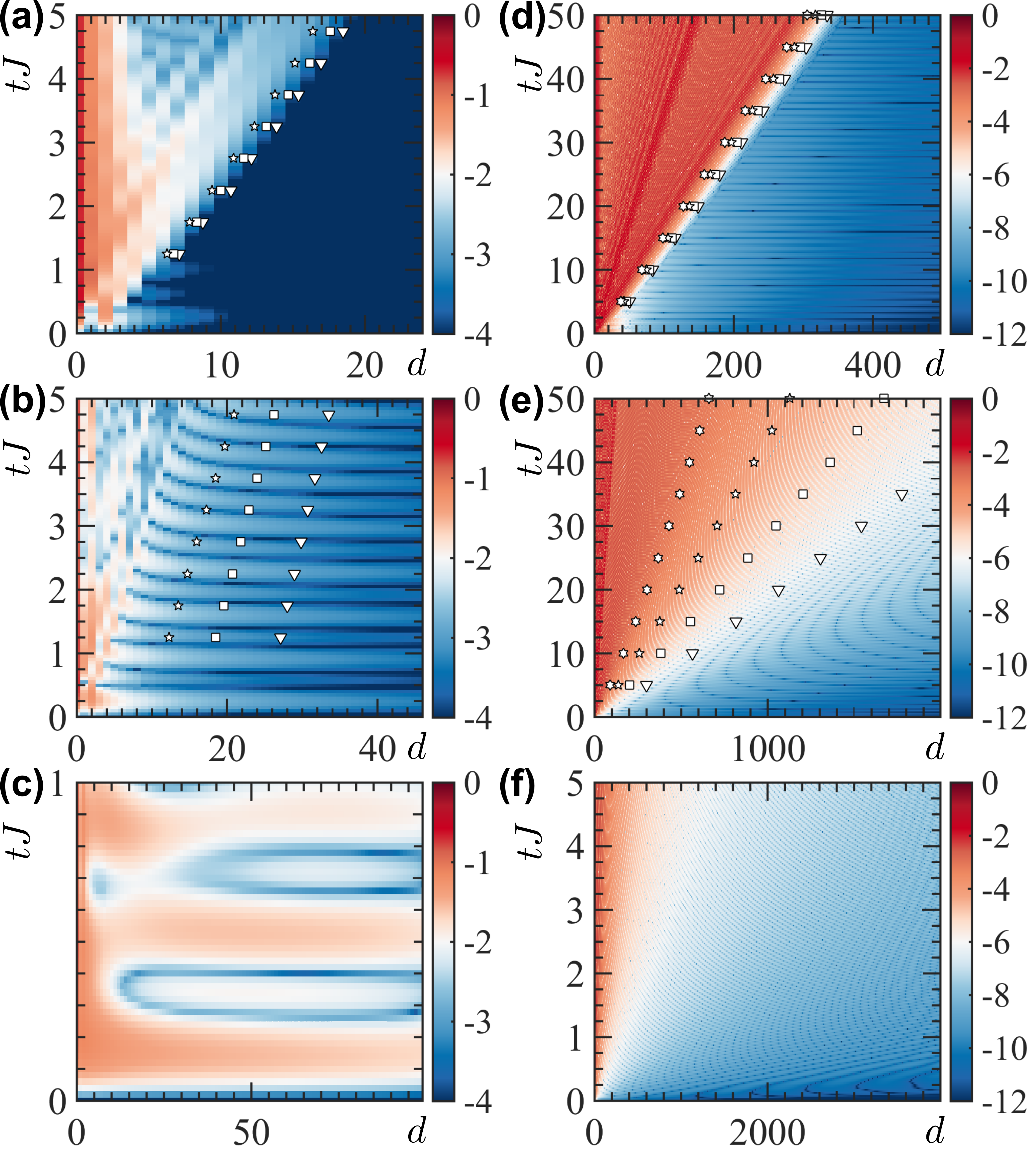}
\par\end{centering}
\caption{\label{fig:comparison_overall_picture} Correlation spreading after global quenches in long-range spin and fermionic models, respectively. (a-c) $\log_{10}|\tilde{C}_d(t)|$ for the LRTI model after the global quench $B/J=\infty \to 1$ is applied. These results are obtained using TDVP approach with MPOs for chains of $M=100$ spins (converged with MPS bond dimension $D=256$). (d-f) $\log_{10}|C_{d}(t)|$ for the LRFH model after the global quench $\Delta/J=10 \to 1$ is applied. These results are obtained using exact numerical computations for $M=10^4$ sites. (a,d) $\alpha=3$,  short-range interactions with a strongly suppressed leakage of correlations outside of the light cone $d/t>v_{\mathrm{gr}}^{\max}$ are observed for both models. Markers indicate contour lines at levels $\log_{10}|\tilde C_d(t)|=[-4,-3\frac{1}{2},-3]$ for the LRTI model and $\log_{10}|C_d(t)|=[-6,-5,-4,-3]$ for the LRFH model. (b,e) $\alpha=3/2$, intermediate-range interactions, the light cone is not sharply defined, but a light-cone effect is observed. Markers indicate averaged contour lines at levels $\log_{10}|\tilde C_d(t)|=[-3\frac{1}{4},-3,-2\frac{3}{4}]$ for the LRTI model and $\log_{10}|C_d(t)|=[-6,-5,-4,-3]$ for LRFH. (c,f) $\alpha=1/2$, no light cone, instant spread of correlations through the entire system. The suppression of correlations at large distances in (f) is discussed in Sec.~\ref{sec:mid-long-range}.}
\end{figure}

\section{Spreading of correlations in time}
\label{sec:correlations}

We begin by analysing the spreading of correlations after a global quench. In order to do this, we look at the spatial mean of the absolute value of characteristic two-site correlation functions for the two systems. For the LRTI system, we choose 
\be
\tilde{C}_d (t) = \left\langle \left|\left\langle \sigma_l^+ (t) \sigma_{l+d}^- (t) \right\rangle \right|\right\rangle _l
\ee
and for the LRFH model, 
\be
C_d\left(t\right)=\left\langle \left|\left\langle c_l^\dagger (t) c_{l+d} (t) \right\rangle \right|\right\rangle_l.
\ee
In these expressions, $\left\langle \ldots\right\rangle _{l}$ indicates the average in space over sites $l$, which depends on whether the boundary conditions are periodic or open.

For quantum systems with finite-range interactions, information is expected to spread with a finite group velocity limited by the Lieb-Robinson bound \cite{Lieb1972}. This forms a light cone for the spreading of information, and the leakage of information outside of this light cone is exponentially suppressed. Such a light-cone-like spreading of correlations was observed for quenches in a Bose-Hubbard model in experiments with a quantum gas microscope \cite{Cheneau2012}. We aim to identify to what extent a behaviour with a linear light cone survives the generalisation to algebraically decaying interactions. 

\subsection{Comparison of the LRTI and LRFH models}

In Fig.~\ref{fig:comparison_overall_picture}, we show examples of correlation spreading after a quantum quench in each of the LRTI (left column of subfigures) and LRFH models (right column of subfigures), comparing the relative behaviour in a regime of short-range interactions where $\alpha>2$ (Figs.~\ref{fig:comparison_overall_picture}(a,d)), 
an intermediate regime for $1<\alpha<2$ (Figs.~\ref{fig:comparison_overall_picture}(b,e)), and a regime of long-range interactions when $\alpha<1$ (Figs.~\ref{fig:comparison_overall_picture}(c,f)). 

In the case where $\alpha>2$, we see a clear linear light cone in the dynamics. Because the light cone is sharp, defining this with threshold values for the correlations leads to a light cone that does not change substantially as the threshold values are changed, as shown in Figs.~\ref{fig:comparison_overall_picture}(a,d). One can see a strongly defined edge with algebraically suppressed correlations outside. 

In an intermediate regime $1<\alpha<2$, the edge of the light cone broadens significantly, and in the case of the LRTI model, the edge is no longer completely linear on the timescales of our calculations. The definition of the edge changes significantly with the chosen threshold value, as can be seen in Figs.~\ref{fig:comparison_overall_picture}(b,e). In the case of the LRFH model we find that the maximum group velocity diverges when $\alpha<2$, which will be discussed in Sec.~\ref{sec:mid-long-range} below.

In the regime of long-range interactions ($\alpha<1$), the spread of correlations becomes even more extreme and light cone effects disappear. Immediately after the quench, correlations start growing over the whole system, and no light cone can be properly defined (note the different scales on the time axis in Fig.~\ref{fig:comparison_overall_picture}(c,f)). Similar behaviour at short to moderate distances is observed for both models. However, for the LRFH model in this regime (Fig.~\ref{fig:comparison_overall_picture}(f)), where we can perform calculations at much longer distances than for the LRTI model, we notice that also in this case the correlations at longer distances are suppressed at short times, despite the long-range interactions. We will discuss this suppression in more detail in Sec.~\ref{sec:mid-long-range} below.

For the LRTI model, qualitatively similar behaviour can be observed also in terms of the mutual information between distant points \cite{Schachenmayer2013}, and has also been seen for local quenches in this model, where an equilibrium state is perturbed by flipping a single spin, as was discussed in Ref.~\cite{Hauke2013}. Although we see quite abrupt changes in the behaviour of the LRTI model at $\alpha=1$ and $\alpha=2$ for system sizes of the order of $M\sim100$ spins, it is difficult to carry out these calculations for longer times and larger systems, and to better delineate these boundaries.  In order to obtain exact results for a global parameter quench, we turn to the LRFH model, for which we will discuss the behaviour for $\alpha>1$ and $\alpha<1$ respectively in the next two sections. 

In Fig.~\ref{fig:logcorrelations}, we plot the correlation function $\log_{10}|C_d(t)|$ as a function of separation distance for several fixed times $tJ$. In the case of short-range interactions, $\alpha>2$, one can clearly see that the connection region between the fast decaying correlation wave and slowly decaying tail occurs over very short distances, as opposed to the case of the intermediate-range correlations $1<\alpha<2$. This leads to a very clearly defined light cone. We also note that for both these regimes ($\alpha>1$) the correlations decay algebraically outside the light cone, as opposed to the exponential decay that would be expected for initially uncorrelated states in models with finite-range or exponentially decaying interactions \cite{Lieb1972}. 

\subsection{Dynamics with short-range interactions $\alpha>2$}
\label{sec:short-range}

\begin{figure}[tb]
\begin{centering}
\includegraphics[width=1\columnwidth]{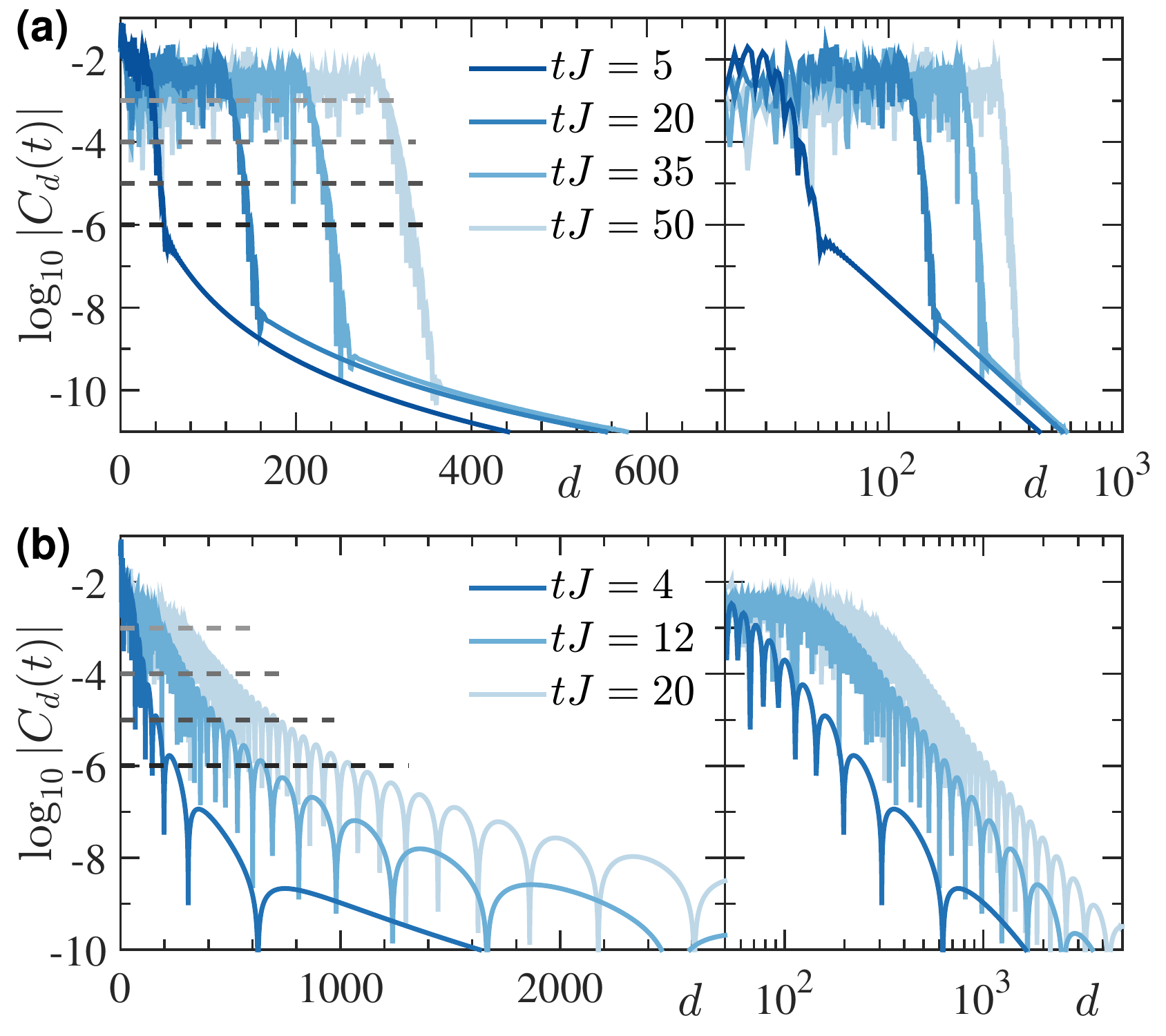}
\par\end{centering}
\caption{\label{fig:logcorrelations} Determining the light cone in the LRFH model. We plot the two-site correlation function $\log_{10}|C_d(t)|$ after global quenches $\Delta/J=10 \to 1$ in the LRFH model with $M=10^4$ sites at different times $tJ$ (a) short-range interactions $\alpha=3$ (b) mid-range interactions $\alpha=3/2$. Dashed lines indicate different threshold levels $\delta=[-6,-5,-4,-3]$ that the correlation function reaches, see the markers in Fig.~\ref{fig:comparison_overall_picture}(d,e). Analogous markers for the LRTI model are in Fig.~\ref{fig:comparison_overall_picture}(a,b). In each case here, on the right hand side, we reproduce the same plots on a double logarithmic scale, showing algebraically decay of correlations outside the light cone as $d\rightarrow\infty$ in both cases.}
\end{figure}

As we already noted above, when $\alpha>2$, the boundary of the light cone is defined clearly (Fig.~\ref{fig:logcorrelations}). We will see below that this result is explained from the behaviour of the density of states in velocity near the maximum group velocity. In this case, we can define a light cone by identifying the position of rapid change in the correlations, essentially equivalently to choosing a threshold value for the correlations, as indicated in Fig.~\ref{fig:logcorrelations}. 

\begin{figure}[tb]
\begin{centering}
\includegraphics[width=1\columnwidth]{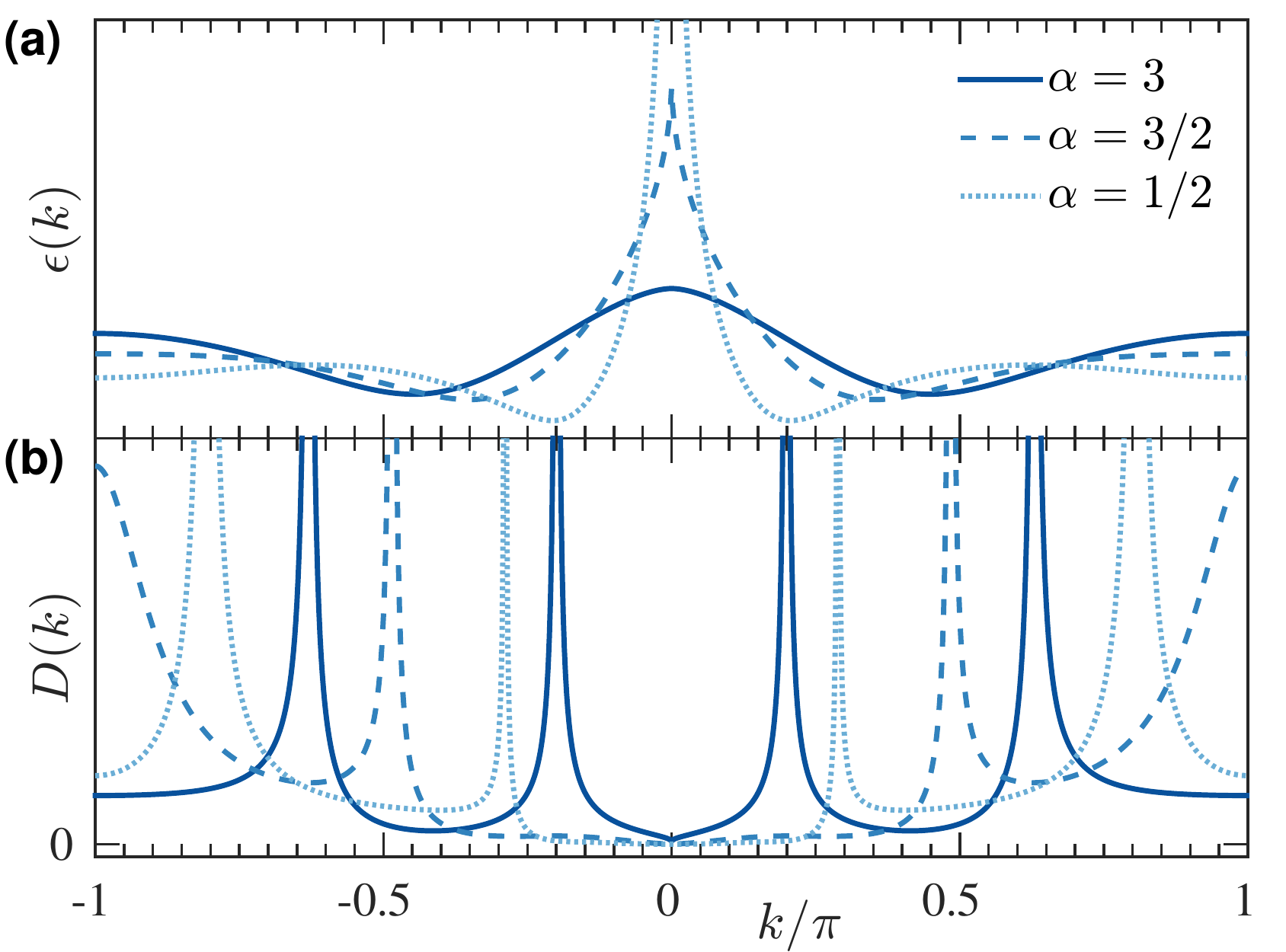}
\par\end{centering}
\caption{\label{fig:dispersion_relation_and_density_of_states} (a) The dispersion relation $\epsilon(k)$ and (b) density of states in velocity $D(k)$ for the LRFH model for $\Delta=J$ and various interaction ranges. In the case of short-range interactions ($\alpha>2$) one can see the smooth behaviour of $\epsilon(k)$ leading to a finite maximum group velocity $v_{\mathrm{gr}}^{\max}=\epsilon'(k^*)$. The corresponding density of states in velocity $D(k^*)$ diverges, which justifies the strong light cone effect. In the case $\alpha<2$, $v_{\mathrm{gr}}^{\max} \propto k^{\alpha-2}$ for $k \to 0$, but density of states in velocity is suppressed, $D(k) \propto k^{3-\alpha}$. As a result there is no domination of infinite velocity excitations and the light cone broadens. In the case $\alpha<1$, the quasiexcitation spectrum become unbounded as well.}
\end{figure}

Alternatively, we can approach the question of the light-cone effect in this regime directly analytically for the LRFH model. To see the general difference between the regime $\alpha>2$ and $\alpha<2$, we can analyse the dispersion relation $\epsilon(k)$, which is plotted in Fig.~\ref{fig:dispersion_relation_and_density_of_states}. We see that for $\alpha>2$ the dispersion as well as its derivative are bounded, i.e.
\begin{equation}
\label{eq:gr_vel_max}
v_{\mathrm{gr}}^{\max}=\max_k \epsilon'(k)<\infty.
\end{equation}

If we look more closely at the time-dependent
part of the Green's function \eqref{eq:greens_function} in the thermodynamic limit
\begin{equation}
F(d,t)=\frac{1}{2}\int_{-\pi}^{\pi}\frac{\mathrm{d}k}{2\pi}\mathrm{e}^{-ikd} \sin \theta_k \sin \delta\theta_k \cos \left( 2\epsilon(k)t \right),
\end{equation}
we see that the corresponding integral has a typical behavior in the ``space-time scaling limit'' \cite{Calabrese2011} of $d,t\rightarrow\infty$, with $u=d/t$ fixed. Using the stationary phase approximation we analytically obtain
\begin{equation}
F(d,t) \approx
\begin{cases}
\mathcal{O} (1/d^{\alpha+2}), & u>v_{\mathrm{gr}}^{\max},\\
\sum_j A_j \cos (B_j),        & u<v_{\mathrm{gr}}^{\max},
\end{cases}
\label{eq:saddle_point_approx}
\end{equation}
where $A_{j}=\left(16\pi\epsilon''\left(k_{j}^{*}\right)t\right)^{-1/2}\sin\theta_{k_j^*} \sin \delta\theta_{k_{j}^{*}}$,
$B_{j}=\pi/4+t\left(k_{j}^{*}u+2\epsilon\left(k_{j}^{*}\right)\right)$,
and $k_{j}^{*}$ are the solutions of the saddle point equation 
\begin{equation}
2\epsilon'\left(k_{j}^{*}\right)+u=0.
\end{equation}
For $u>v_{\mathrm{gr}}^{\max}$, the usual argument for the exponential decay of the integral due to the lack of a stationary point fails due to a non-analyticity at $k=0$. In this case, using $2\pi$-periodicity of the integrand, its non-analytical part was extracted. The contribution of this part decays only algebraically in a distance $d$ and prevails over the exponentially small contribution of the remaining analytical part. Thus, we obtain the power law exponent of the correlation decay outside of the light cone (see Eq.~\eqref{eq:saddle_point_approx}). 

On the other hand, for $u<v_{\mathrm{gr}}^{\max}$, the existing saddle point gives the major contribution to the integral. As a result, the correlations inside of the light cone decay slowly in time, $\mathcal{O} (t^{-1/2})$, which corresponds to Fick's law of diffusion. We find that both approximations agree well with full numerical solutions up to finite size corrections.

At the same time, in Fig.~\ref{fig:comparison_overall_picture}, and through the following analysis of the density of states in velocity, we see that the light cone effect is strong only when $\alpha>2$. To understand this we consider both the dispersion relation and the density of states in velocity as a function of the wave vector (Fig.~\ref{fig:dispersion_relation_and_density_of_states}),
\begin{equation}
D\left(k\right)=\frac{M}{\pi}\left|\frac{\mathrm{d}v_{\mathrm{gr}}(k)}{\mathrm{d}k}\right|^{-1}=\frac{M}{\pi}\left|\frac{\mathrm{d}^{2}\epsilon(k)}{\mathrm{d}k^{2}}\right|^{-1}.
\end{equation}
We see that for $\alpha>2$, the density of states in velocity diverges exactly at the value of $v_{\mathrm{gr}}=v_{\mathrm{gr}}^{\mathrm{max}}$, which means that depending on the type of quench (the final occupation number) it is possible to excite infinitely many quasi-particles propagating with the maximum group velocity. Those quasi-particles propagating through the system will form the well-defined front of correlations. And as one can see in Fig.~\ref{fig:dispersion_relation_and_density_of_states}(b) the maximum group velocity is finite in this regime, $v_{\mathrm{gr}}^{\mathrm{max}}<\infty$. 

\subsection{Dynamics with intermediate and long-range interactions  ($\alpha<2$) in the LRFH model}
\label{sec:mid-long-range}

The situation changes when a kink appears at $k=0$ in the dispersion relation for $1<\alpha<2$
\begin{equation}
\epsilon(k) \propto \epsilon_0 + c k^{\alpha-1},
\end{equation}
in this regime the dispersion is bounded, but the velocity diverges. For $\alpha<1$ even the dispersion diverges for small momenta $k$
\begin{equation}
\epsilon(k) \propto k^{\alpha-1},\,\,\, k\rightarrow0.
\end{equation}
Hence, for $\alpha<2$ there no longer is a finite maximum
group velocity and for $\alpha<1$ the spectrum of quasi-excitation become unbounded. This result can be seen analytically in the case of open boundary condition, as the result of the polylogarithm behaviour in Eq.~\eqref{eq:Jk_OBC} for $\alpha<2$. As a result the light cone boundary becomes washed out for these regimes.

Even though the group velocity diverges for $\alpha<2$, the density of states in velocity around $k=0$ is suppressed, $D(k)\propto k^{3-\alpha}$. The combination of both these facts means that although the correlations can to some extent build instantly through the entire system, they grow very slowly, $v_{\mathrm{gr}}\left(k\right)D\left(k\right)\propto k\rightarrow0$ for $k \to 0$. Because of this suppression the correlation spread for fermions in Fig.~\ref{fig:comparison_overall_picture}(f) does not seem as immediate as in the case of spins in Fig.~\ref{fig:comparison_overall_picture}(c), which will be discussed in Sec.~\ref{sec:HPapproximation}.

In terms of the saddle point approximation, the situation is very different from the case of $\alpha>2$, but the approximation is still very good. Now there is always at least one saddle point, because the group velocity is unbounded, $v_{\mathrm{gr}}^{\max}=\infty$. Therefore we have only the second part of the solution of Eq.~\eqref{eq:saddle_point_approx}. 

In order to understand the behaviour in the case of small $\alpha$ it is useful to consider the contribution of the smallest saddle point. In particular we focus on $d \gg ut$, in which case this saddle point $k^*$ occurs close to zero. We can work out the result of the saddle point approximation to be 
\begin{equation}
F(d,t)\propto\frac{t^{3/2}}{d^2}
\begin{cases}
\left(t/d\right)^{\frac{3-\alpha}{4-2\alpha}}, & \alpha<1\\
\left(t/d\right)^{\frac{3\alpha-1}{4-2\alpha}}, & 1<\alpha<2
\end{cases}.
\end{equation}
Hence, although there is no light cone, the time dependent part of the correlation function is small and scales like a power-law in time.

\subsection{Holstein-Primakoff approximation for the LRTI model}
\label{sec:HPapproximation}

In order to get a more detailed picture of the dynamics in the LRTI model at short times, we consider an analytical treatment in the Holstein-Primakoff approximation. This relies on the initial ordering of the spins along the $z$-axis (from the initial state we chose previously), and should be a good approximation as long as this order remains. We note that this is expected to be a better approximation for non-zero values of the transverse field $B$, as this supports retention of the ordering.

Considering this initial state with $\left|\psi_{0}\right\rangle =\otimes_{l}\left|\downarrow\right\rangle _{l}$, and writing $\hbar\equiv 1$, we can rewrite Eq.~\eqref{eq:H_LRTI} using the transformation 
\[
\begin{cases}
S_{l}^{z}=\tilde a_{l}^{\dagger}\tilde a_{l}-S\\
S_{l}^{-}\equiv S_{l}^{x}-iS_{l}^{y}=\sqrt{2S-\tilde a_{l}^{\dagger}\tilde a_{l}}\tilde a_{l}\\
S_{l}^{+}\equiv S_{l}^{x}+iS_{l}^{y}=\tilde a_{l}^{\dagger}\sqrt{2S-\tilde a_{l}^{\dagger}\tilde a_{l}}
\end{cases},
\]
where $S=\sigma/2$ are spin operators for spin-1/2 and $\tilde a_{l}$ are bosonic annihilation operators. Starting from the initially fully polarized state we consider dynamics only on timescales when the initial order is still preserved, i.e. $\bigl\langle \tilde a_{l}^{\dagger}\tilde a_{l} \bigr\rangle /2S\ll1$,
then we can take
\[
\begin{cases}
S_{l}^{z}\approx \tilde a_{l}^{\dagger}\tilde a_{l}-S\\
S_{l}^{-}\approx \sqrt{2S}\tilde a_{l}\\
S_{l}^{+}\approx \sqrt{2S}\tilde a_{l}^{\dagger}
\end{cases}.
\]
Up to a constant shift, we can then write the LRTI model as 
\begin{eqnarray*}
H_{\rm HP} &= & S\sum_{l\neq j}J_{lj}\left(\tilde a_l \tilde a_j + \tilde a_l^\dagger \tilde a_j+\mathrm{h.c.}\right)+2B\sum_l \tilde a_l^\dagger \tilde a_l.\label{eq:H_HPform}
\end{eqnarray*}
Taking the Fourier transformation to momentum space operators,
\[a_{k}=\frac{1}{\sqrt{M}}\sum_{l=1}^{M}\mathrm{e}^{-i\frac{2\pi kl}{M}}\tilde a_{l}
\]
we can then rewrite the Hamiltonian as 
\begin{eqnarray}
H_{\rm HP} &=& 2 \sum_{k=0}^{M/2-1} \left[ S\mathcal{J}(k) \left( a_{M-k} a_k + a_k^\dagger a_{M-k}^\dagger \right) \right. \nonumber \\
& & \left. + \left( S\mathcal{J}(k)+B \right) \left( a_k^\dagger a_k + a_{M-k} a_{M-k}^\dagger \right) \right],
\label{eq:H_HP_k}
\end{eqnarray}
where $\mathcal{J}(k)$ is defined in Eq.~\eqref{eq:Jk_OBC}. Then via the Bogoliubov transformations for the bosonic field
\begin{eqnarray*}
\left(\begin{array}{c}
\gamma_{k}\\
\gamma_{M-k}^{\dagger}
\end{array}\right)=\left(\begin{array}{cc}
u & v\\
v^{*} & u^{*}
\end{array}\right)\left(\begin{array}{c}
a_{k}\\
a_{M-k}^{\dagger}
\end{array}\right),
\end{eqnarray*}
where $u=\cosh\left(\phi_{k}/2\right)$, $v=\sinh\left(\phi_{k}/2\right)$. Eq.~\eqref{eq:H_HP_k} can be diagonalized if
\[
\mathrm{e}^{2\phi_{k}}=1+\frac{2S\mathcal{J}\left(k\right)}{B}.
\]
Then the dispersion relation reads
\begin{equation}
\epsilon(k)=2B\sqrt{1+\frac{2S\mathcal{J}\left(k\right)}{B}} .\label{eq:dispersion_magnons}
\end{equation}

Note that in order for $\epsilon(k)$ to be real, we require the expression under the root to be real. Then, using Eq.~\eqref{eq:Jk_OBC}, we obtain the following constraint
\begin{equation*}
\left|B/J\right| \geq -2 \, \mathrm{sign} (B/J) \mathrm{Li}_\alpha (-\mathrm{sign}(B/J)),
\end{equation*}
which sets the limits for $B$ values such that the Holstein-Primakoff transformation is stable for all modes $k$. 

Using the limiting behaviour of the polylogarithm, we then obtain the divergence of the dispersion relation for $\alpha<1$,
\begin{equation}
\epsilon(k)\propto k^{\frac{\alpha-1}{2}},\,\left(\alpha<1,\, k\rightarrow0\right)
\label{eq:disp_HP_limit}.
\end{equation}
In general the dispersion relation and density of states in velocity have the same featured as for the LRFH model in Fig.~\ref{fig:dispersion_relation_and_density_of_states}. 

We observe a similar divergence in the Bogoliubov angle $\delta\phi_k=\phi_k-\phi_k^0$, where $\phi_k^0$ is the pre-quenched Bogoliubov angle for the Hamiltonian with the field $B_0=\infty$ instead of $B$. When $\alpha<1$, modes near $k=0$ will dominate the spread of entanglement, and we will observe a transition of behaviour with respect to $\alpha>1$, when the dispersion relation is regular for $k=0$. This explains the transition in behaviour that we observed at $\alpha=1$ numerically, as outlined in Sec.~\ref{sec:transitiondiscussion}.

Analogous to the other models we consider the correlation matrix for the bosonic particles
\be
\hat{C}_d (t) = \left\langle \left| \left\langle a_l^\dagger(t) a_{l+d}(t) \right\rangle \right|\right\rangle_l.
\ee
after a global quench of the transverse field.

\begin{figure}[tb]
\begin{centering}
\includegraphics[width=1\columnwidth]{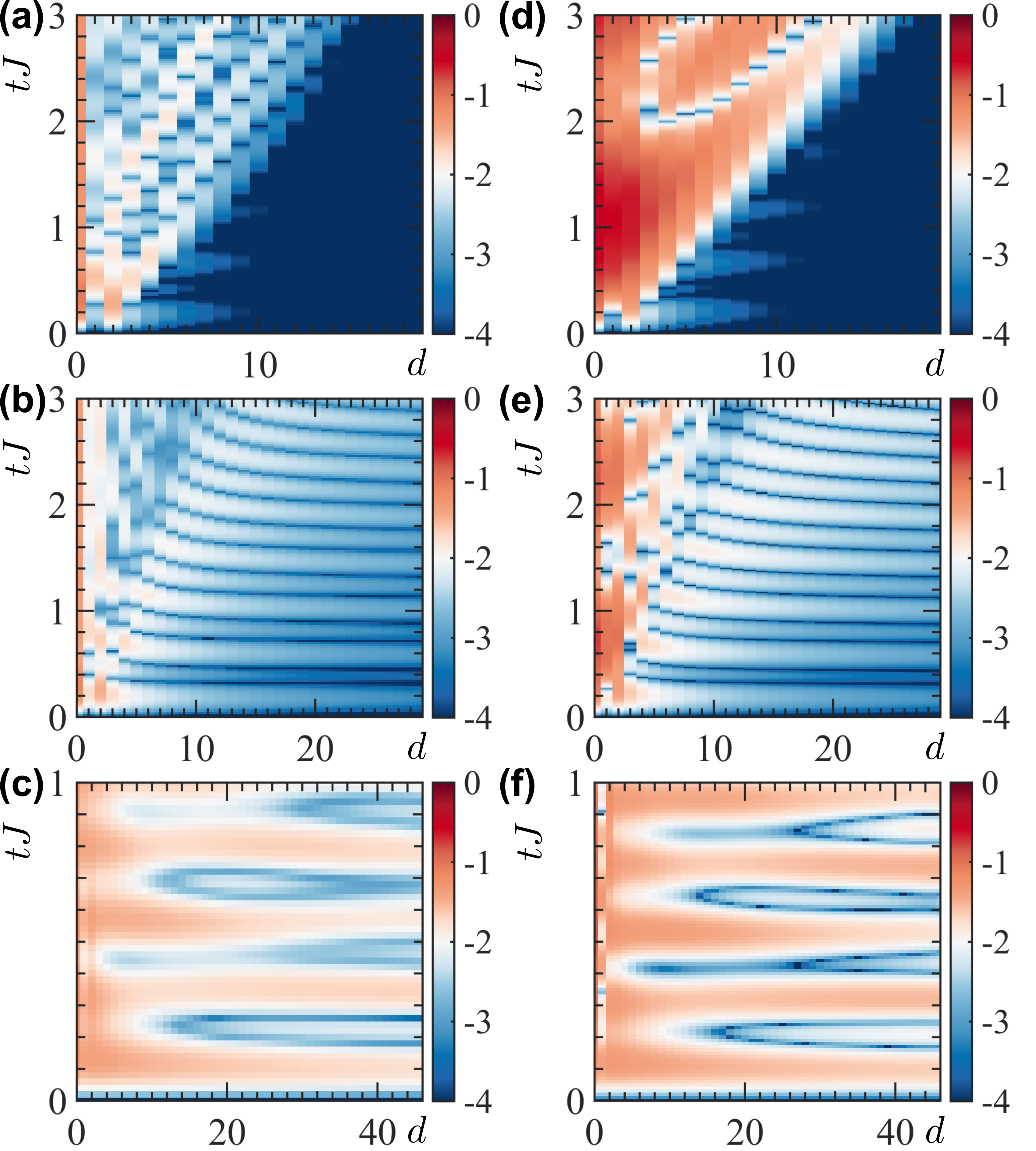}
\par\end{centering}
\caption{\label{fig:comparison_TI_and_HP_approx} Correlation spreading in the LRTI model after the global quench $B/J=\infty \to 2$ is applied in the system of $M=100$ spins. (a-c) $\log_{10}|\tilde{C}_{d}(t)|$ calculated numerically via MPS/MPO methods (as in Fig.~\ref{fig:comparison_overall_picture}(a-c)), and (d-f) $\log_{10}|\hat{C}_{d}(t)|$ calculated analytically via the Holstein-Primakoff approximation. (a,d) $\alpha=3$, short-range interactions case. (b,e) $\alpha=3/2$, intermediate-range interactions. (c,f) $\alpha=1/2$, long-range interactions.}
\end{figure}

In Fig.~\ref{fig:comparison_TI_and_HP_approx} we can see the comparison of correlation functions for the LRTI model calculated with MPS/MPO methods and via the Holstein-Primakoff transformation after the system is quenched from the fully polarized state. The initial order of the state is preserved only at times of a fraction of $tJ$ for moderate fields $B$, which does not allow us to use this approximation for quantitative analysis for longer times. By increasing the absolute value of the field one can extend the lifetime of the order, and the Holstein-Primakoff transformation will be valid for longer times. At the same time, the exchange term becomes stronger for longer range interactions and the initial spin order lasts for shorter times. It should be noted, although the quantitative agreement lasts for short times, key qualitative aspects of the dynamics are typically captured over longer timescales.

\section{Entanglement Growth}
\label{sec:entanglement}

In this section we quantify the time-dependence of entanglement in spatial modes of the system. In order to do this, we can split the system into two parts. These could be a part in the centre, and the rest of the system bulk, or a single bipartite splitting of the 1D system (e.g., the half-chain splitting directly in the centre of the system). If we refer to these two parts of the system as $A$ and $B$, then assuming the quantum state of the whole system to be pure, if it cannot be written as a product of states of both parts, i.e. $\left|\psi\right\rangle \neq\left|\psi_{A}\right\rangle \otimes \left|\psi_{B}\right\rangle $, then $A$ and $B$ are entangled. As a measure of spatial entanglement
we use the von Neumann entropy of the reduced density matrix for one part of the system, $\rho_{A}=\mathrm{tr_{B}}\left(\left|\psi\right\rangle \left\langle \psi\right|\right)$, given by
\begin{equation}
S_{\mathrm{vN}}=-\mathrm{tr}\left(\rho_{A}\log_{2}\rho_{A}\right).
\end{equation}
In the case of local interactions (or interactions that exponentially decay with distance), the bipartite entanglement entropy grows linearly in time after a quench \cite{Calabrese2005,Fagotti2008}. This is related to a finite speed of excitations propagating in the system, which leads to a linear rate of quantum information exchange between partitions and linear growth of entanglement. We can then naturally ask to what extent the behaviour that is known for short-range interactions persists for the models we consider here.

Below we first consider the growth of entanglement for the LRTI model, which we again compute using matrix product operator techniques. We then analyse analytically the entanglement growth in the LRFH model. In each case, we find that the behaviour mostly follows the same form as for short range interactions. The counter-intuitive exception to this is that when the initial state is reasonably symmetric, for long-range interactions the short to medium time entanglement growth can be very sublinear --- i.e., in certain regimes having long-range interactions actually suppresses the growth of entanglement relative to short-range interactions.

\subsection{Entanglement growth in the long-range transverse Ising model}

\begin{figure}[tb]
\begin{centering}
\includegraphics[width=1\columnwidth]{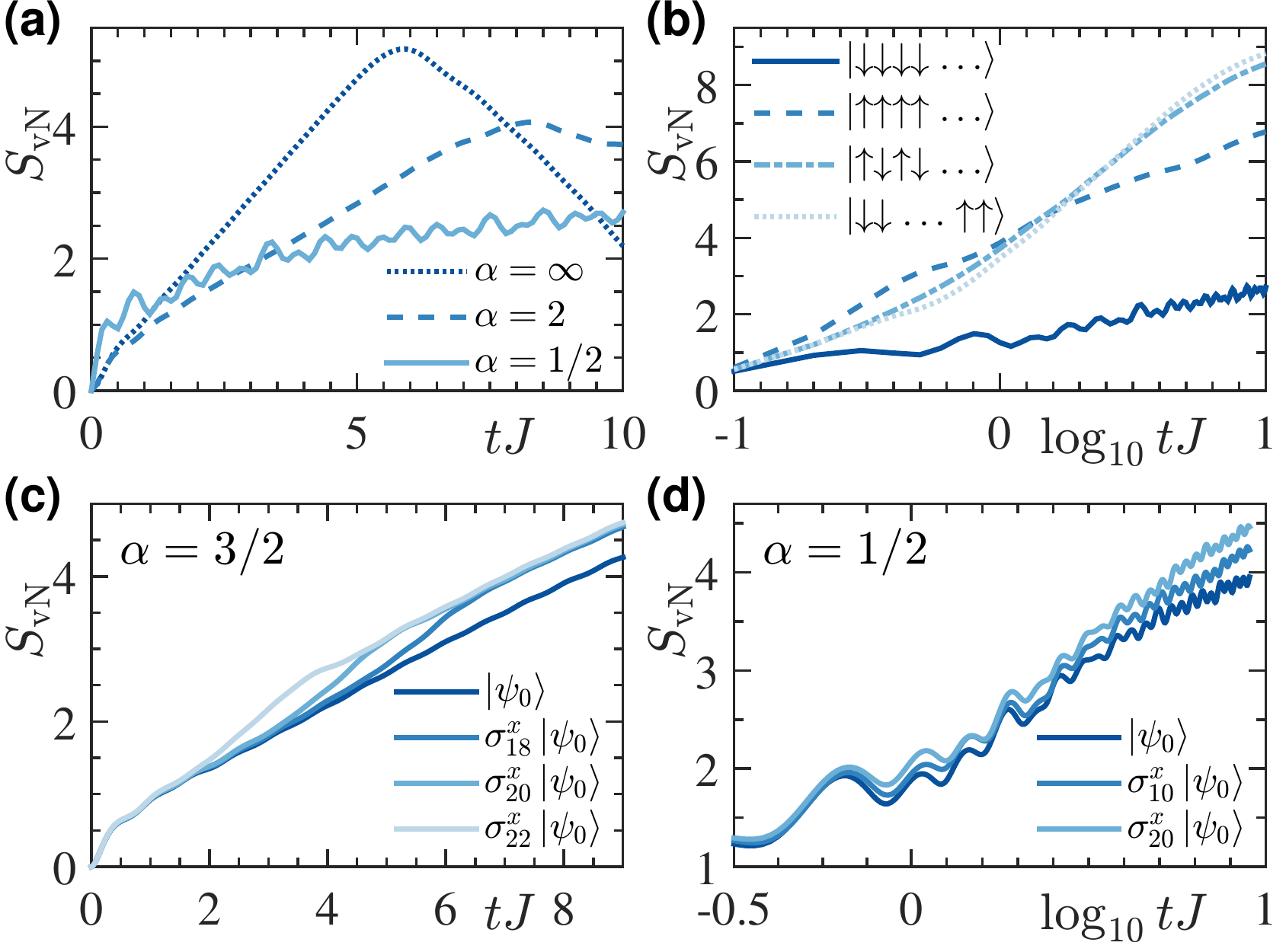}
\par\end{centering}
\caption{\label{fig:entanglementLRTI} Bipartite entanglement growth after the global quench in LRTI model with open boundary conditions.
(a) Half-chain entanglement entropy as a function of time for $M=20$ spins beginning in the fully polarised state $\left| \psi_0 \right\rangle=\left|\downarrow\downarrow\downarrow\ldots\right\rangle$ along the axis of the magnetic field $B=J$ (from exact diagonalisation). (b) The same as (a), but for $\alpha=1/2$, and starting from a selection of initial states (2 fully polarised states in the opposite directions, N\'eel ordered state, and a product state with spins down in the left half of the chain and up in the right half). (c) Linear and (d) sublinear growth of entanglement entropy (shown on a logarithmic scale) for the LRTI model and selection of initial states, now for $M=50$ spins, computed with MPS methods (converged with MPS bond dimension $D=256$). }

\end{figure}

In Fig.~\ref{fig:entanglementLRTI}, we plot representative calculations for the growth of entanglement entropy in the LRTI model, beginning from a selection of different initial states. In Fig.~\ref{fig:entanglementLRTI}(a) we see clearly the change in characteristic behaviour as we go from the nearest neighbour interaction limit ($\alpha\rightarrow\infty$) to long-range interactions, beginning in a fully polarised state with all spins down. 

In Fig.~\ref{fig:entanglementLRTI}(b), we focus on the case where $\alpha<1$, and identify clearly sublinear behaviour in the entanglement entropy growth at short times. This is approximately linear on a logarithmic scale, but depends strongly on the structure of the initial state, with much slower growth at intermediate times for the initially fully polarised state, compared with other spin configurations or with short-range interactions. The intuitive explanation for this restriction of entanglement growth at intermediate times is that the structure of the initial state, when combined with the symmetry of the Hamiltonian, prevents the system from accessing large sections of the Hilbert space at short to intermediate times, as it becomes somewhat stuck in its initial symmetry sector. The ultimate limit of this occurs when $\alpha\rightarrow 0$, where the Hamiltonian is fully symmetric. In that case, if we begin with a completely polarized spin state, the system will be at all times restricted to completely symmetric spin states, substantially limiting the maximum entanglement entropy that can be reached \cite{Schachenmayer2013}. 

In Figs.~\ref{fig:entanglementLRTI}(c,d) we consider initial states with a single spin flipped compared with the fully polarised state. We see that both the linear behaviour for $\alpha>1$, and the sublinear behaviour for $\alpha<1$ are quite robust to small changes in the initial state like this at short times.  At the same time, a small difference between the two cases can be observed, because in calculating the half-chain entanglement entropy, we notice that the changes in the dynamics induced by the spin flips take time to propagate across the divide between the different parts of the system and affect the half-chain entanglement entropy for $\alpha>1$. This can be seen from the equality of all curves at short times in Fig.~\ref{fig:entanglementLRTI}(c). In contrast, for $\alpha<1$, the disturbance in the initial state has immediate effect on entanglement independent of its position in the chain, as seen in Fig.~\ref{fig:entanglementLRTI}(d).

\subsubsection{Behaviour in limiting cases}

Before treating the variation in behaviour with changing $\alpha$ in general, it is worthwhile to consider the limiting cases of very short or very long-range interactions. In the case of the nearest-neighbour interactions ($\alpha\rightarrow\infty$) the transverse Ising model can be studied analytically and the dynamics after global quantum quenches has been considered in numerous works, see e.g.  \cite{Igloi2000, Rossini2009, Rossini2010, Calabrese2011, Rieger2011, Calabrese2012a, Calabrese2012b, Schuricht2012, Foini2012, Essler2012, Fagotti2013, Caux2013}. In this limit the Ising chain can be mapped onto a model of free fermions, which has the twofold degenerate dispersion relation $\epsilon(k)=2\sqrt{(J-B)^2+4JB\sin^2(k\lambda/2)}$ for $k\neq0$, where $\lambda$ is the distance between spins. The fastest quasi particles in this model move at the Lieb-Robinson velocity $v_{\mathrm{LR}}=v_{\mathrm{gr}}^{\max}=2\lambda J$ for $B \geq J$ and $v_{\mathrm{LR}}=v_{\mathrm{gr}}^{\max}=2\lambda B$ for $B<J$. 
By performing the global quench of the system parameters we excite counter-propagating entangled pairs of quasi particles at certain points of the system. The spatial entanglement between two parts of the system grows as one quasi particle of a pair crosses the border. Since the pairs are spreading with a finite maximum speed the spatial entanglement entropy is limited by the Lieb-Robinson bound, which leads to a linear growth of the entanglement entropy $S_{\mathrm{vN}} \leq C_1 v_{\mathrm{LR}}t+C_2$, where $C_1$ and $C_2$ are constants, as is discussed in more detail in Sec.~\ref{sec:entanglementanalytical}.

In the opposite case of all-to-all interactions with $\alpha=0$, the model can also be analytically solved via mapping with the Lipkin-Meshkov-Glick model, for which entanglement properties have been studied \cite{Vidal2004,Wilms2012}. By starting from a fully polarised initial state with spins aligned with the external field $B$, the dynamics is fully restricted to the subspace of Dicke states with finite well defined maximum spatial entanglement \cite{Latorre2005}. Hence, the entanglement entropy will always be restricted $S_{\mathrm{vN}} \leq \log_2 (M/2+1)$, where $M$ is the number of original spins, intuitively explaining the slow and bounded growth of entanglement with respect to the case of long-range interactions.

\subsubsection{Transition at $\alpha=1$}
\label{sec:transitiondiscussion}
In order to understand the change in behaviour that occurs at $\alpha=1$, we need to go well beyond the intuitive discussion above for $\alpha \rightarrow 0$. In Sec.~\ref{sec:entanglementanalytical}, we will show that the analogous transition in the LRFH model occurs as a result of changes in the dispersion relation, and show how this directly relates to qualitative changes in the growth of bipartite entanglement.

Although the LRTI model is not analytically solvable, as we showed in Sec.~\ref{sec:HPapproximation}, we can treat this model for initial states with all of the spins aligned along the $z$-axis in the Holstein-Primakoff approximation, which gives an exactly solvable model that is relevant in the description of short to intermediate time dynamics when the ordering in the alignment of the spins is still present. This allows us to identify the change in qualitative behaviour at $\alpha=1$. Within that approximation, we observed (also as shown in Sec.~\ref{sec:HPapproximation}) that for $\alpha<1$, there is a divergence in the excitation spectrum, and hence in the group velocity, as $k\rightarrow 0$. For initial states where the quench excites significant excitations near $k=0$, the dynamics of these quasiparticles then dominate the growth of entanglement, changing the entanglement entropy growth away from linear. This is completely analogous to the case of the LRFH model, which we will now discuss.

\subsection{Entanglement growth in the long-range fermionic hopping model}

In this section we discuss analytical forms for the growth of the entanglement in the LRFH model, which we also evaluate numerically. We will see the same qualitative behaviour as we saw for the LRTI model: linear growth for all $\alpha$ for generic initial states, but a marked change in behaviour for particular initial states and long-range interactions. We can understand this analytically from the behaviour of and contribution to the entanglement growth from the quasiparticles with $k\rightarrow 0$.  

\subsubsection{Quench from $\Delta =4J$ to $\Delta=J/5$}
\label{sec:entanglementanalytical}

\begin{figure}[tbp]
\includegraphics[width=\columnwidth]{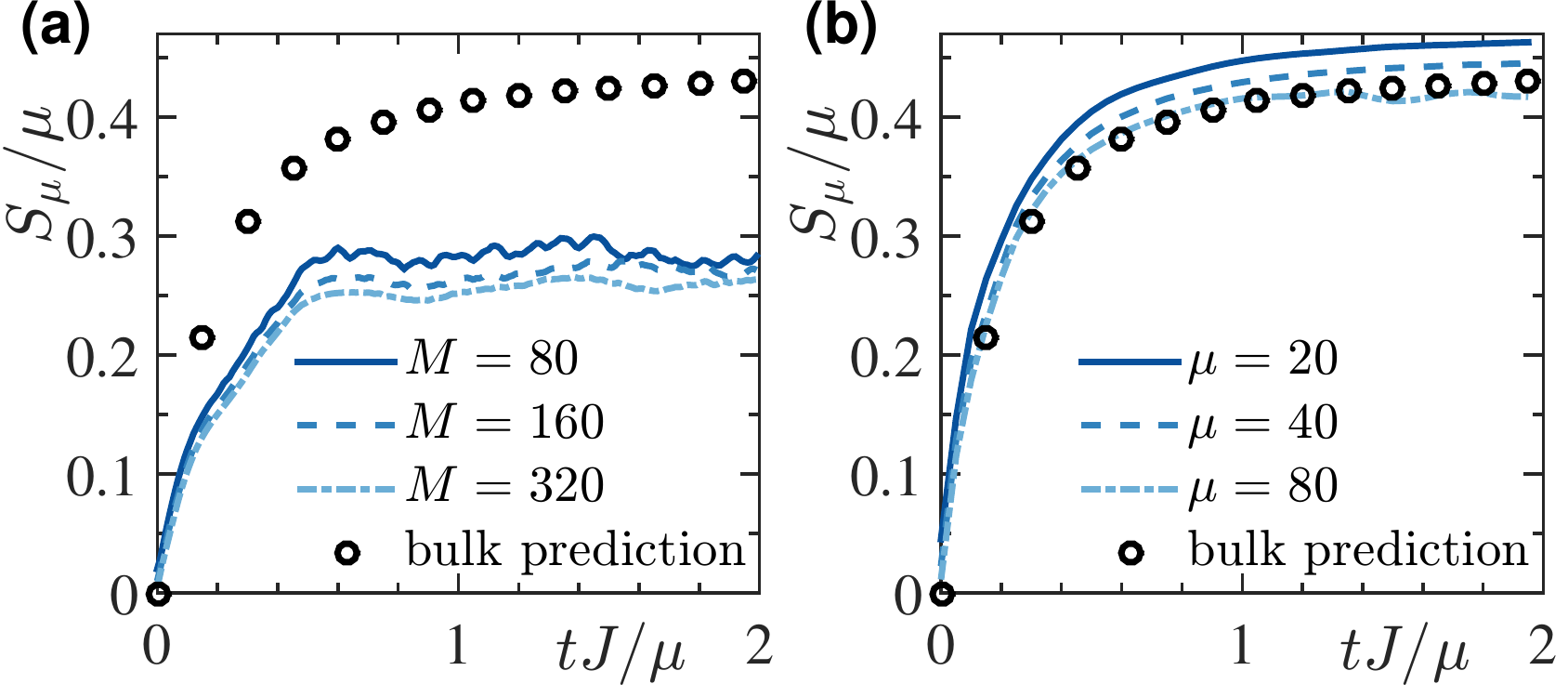}\\
\caption{
The entanglement entropy in LRFH model with $\alpha=0.8$, computed using exact numerical techniques for various system sizes $M$ and subsystem sizes $\mu$ after the global quench of $\Delta/J= 4 \to 1/5$. (a) Entanglement dynamics for the equal partition case with $\mu=M/2$. The circular markers show the limit described by~\eqref{eq:S} for large subsystems in the bulk (i.e., $\mu, M \rightarrow \infty$ and $\mu\ll M$). 
(b) Entanglement dynamics for a selection of system and subsystem sizes, which is shown to converge to the prediction \eqref{eq:S} with increasing $\mu$ for $\mu \ll M = 640$. The circular markers again indicate \eqref{eq:S} for large subsystems in the bulk.
\label{fig:svnFermions1}
}
\end{figure}

We consider the entanglement entropy as a function of time, both for a bipartite splitting in the centre of the system, and the entanglement of blocks of size $\mu$ with the remainder of the system (especially in the thermodynamic limit where $M\rightarrow \infty$). In Fig.~\ref{fig:svnFermions1} we show numerical results for the growth of the entanglement entropy in a quench within the LRFH model from $\Delta=4J$ to $\Delta= J/5$. Fig.~\ref{fig:svnFermions1}(a) shows the half-chain entanglement entropy $\mu=M/2$, and in Fig.~\ref{fig:svnFermions1}(b), we consider a small chain $\mu\ll M$. In each case, we see a fairly typical generic form for the entanglement growth, in which it saturates at the value and time that is proportional to the size of the subsystem, $\mu$. It should be noted that the half-chain entropy per unit length, is generally smaller than for smaller subsystems $\mu$.

It is useful to co compute an analytical expression for this growth in the thermodynamic limit, where $\mu\ll M \rightarrow \infty$. Here we can even consider open boundary conditions as the opposite to the derivations in Sec.~\ref{sec:LRFH_PBC_derivation}. In this limit, it is convenient to define the Fourier transform of the long range term (analogous to Eq.~\eqref{eq:Jk_PBC})
\begin{eqnarray}
\label{eq:Jk_OBC}
\mathcal J(k)&=&J \sum_{d=-\infty\atop d\neq 0}^{\infty}\frac{e^{-i d k}}{|d|^\alpha}=J\Bigl[{\rm Li}_\alpha(e^{i k})+{\rm Li}_\alpha(e^{-i k})\Bigr] \nonumber \\
&=&2J\mathrm{Re}[{\rm Li}_\alpha(e^{i k})]\, 
\end{eqnarray}
where ${\rm Li}_n(z)$ is the polylogarithm of order $n$ and argument $z$. In terms of this, we get the same dispersion relation and the Bogoliubov angle as in Sec.~\ref{sec:LRFH_PBC_derivation} but with $\mathcal J(k)$ instead of $\bar {\mathcal J}(k)$.

If we consider pre-quench Hamiltonians that are reflection symmetric and can be diagonalised by the Bogoliubov transformation, then after the quench the entanglement entropy per unit length of large subsystems far away from the boundaries (i.e., in the thermodynamic limit) is given by~\cite{Fagotti2008,Calabrese2012a}
\be\label{eq:S}
\frac{S_\mu}{\mu}\sim\int_0^\pi\frac{\mathrm d k}{\pi} \min\Bigl(1,2 \left| \varepsilon'(k) \right| \frac{t}{\mu}\Bigr)G\left( \cos \delta\theta_k \right) ,
\ee
where
\be
G(x)=-\frac{1+x}{2}\log \frac{1+x}{2}-\frac{1-x}{2}\log \frac{1-x}{2},
\ee
and $\delta\theta_k$ is the difference of the Bogoliubov angles before and after the quench, as in Eq.~\eqref{eq:deltatheta}. We plot this expression in Fig.~\ref{fig:svnFermions1}, and see that it fits very well for larger subsystems in Fig.~\ref{fig:svnFermions1}(b).

\begin{figure}[tbp]
\includegraphics[width=\columnwidth]{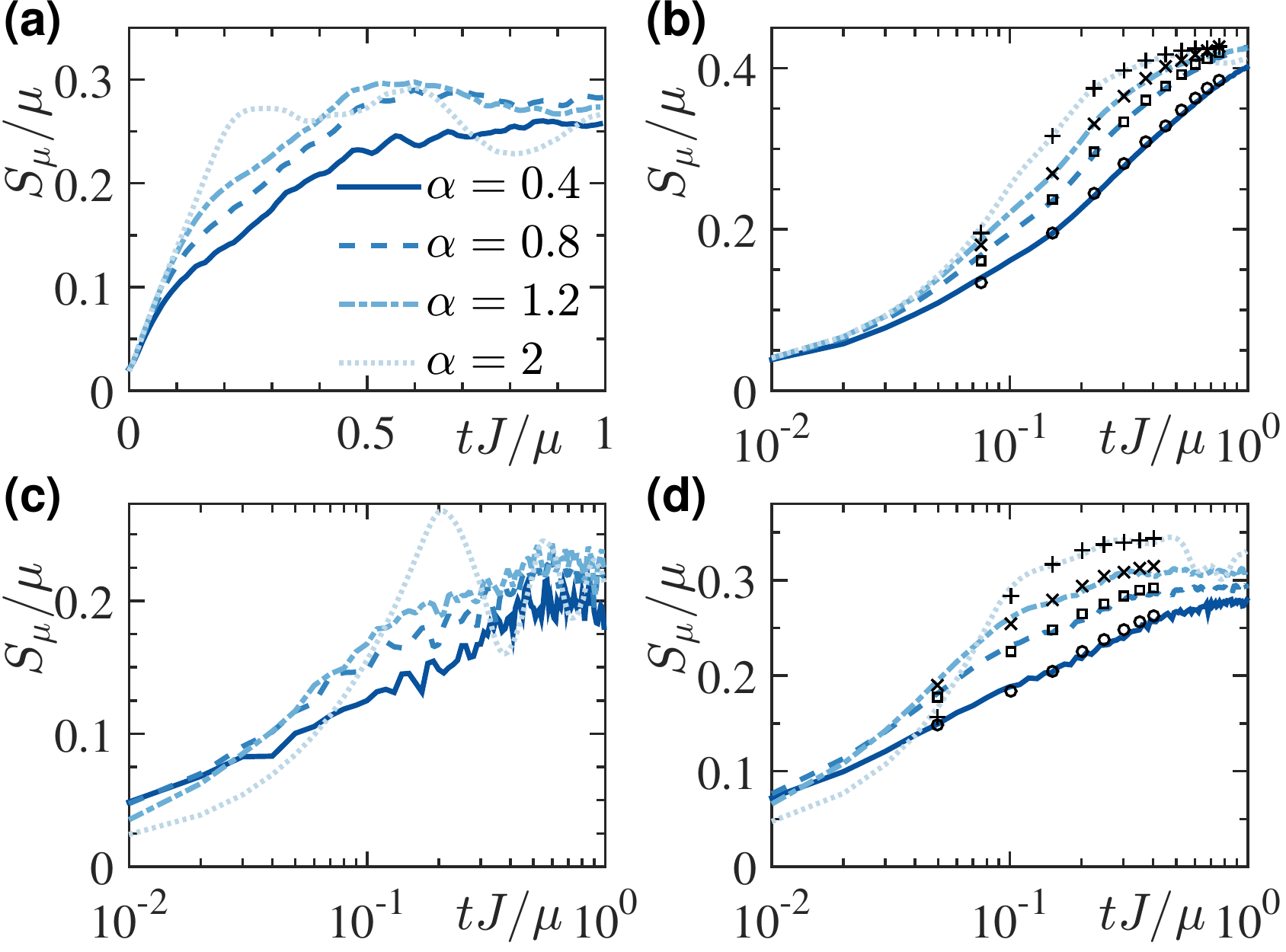}\\
\caption{
The entanglement entropy in LRFH model computed using exact numerical techniques with a selection of interaction range exponents $\alpha$, showing results for $\Delta=J/5$, and two different initial states two cases: (a,b) Beginning from the ground state for $\Delta=4J$, and (c,d) beginning from the ground state of a critical Ising model, as detailed in the text.
(a,c) The half-chain entanglement entropy $\mu=M/2=40$, and (b,d) the subsystem entanglement entropy for $\mu=M/8=40$. Markers denote the bulk prediction from \eqref{eq:S}.
(a) For $\alpha\geq 2$ there is a time window of clean linearity. At short times the leading behaviour seems to be linear, but the subleading terms (in $tJ/\mu$) are also affected by the divergence of the dispersion relation.
(b) The agreement with the bulk prediction is excellent, but at sufficiently large $tJ/\mu$ the discrepancy due to the finiteness of $M$ is clearly visible, especially for the largest values of $\alpha$. 
(c) There are evident differences with respect to (a) due to the fact that quasiparticles with momentum close to zero turn out to give a finite contribution to the entropy. While for $\alpha\geq 2$ the entropy seems to grow linearly, for small $\alpha$ the leading contribution at short time growth logarithmically in time.
(d) The agreement with prediction~\eqref{eq:S} is still good, although the leading correction seems to be more complicated than a simple constant. 
\label{fig:svnFermions2}
}
\end{figure}

We see that Eq.~\eqref{eq:S} can be always bounded by the velocity term
\begin{eqnarray}
\label{eq:S_bound}
& &\int_0^\pi\frac{\mathrm d k}{\pi} \min \left(1,2 \left| \varepsilon'(k) \right| \frac{t}{\mu} \right) G \left( \cos  \delta\theta_k \right) \nonumber  \\
&\leq& 2 \frac{t}{\mu} \int_0^\pi\frac{\mathrm d k}{\pi} |\varepsilon'(k)| G \left( \cos  \delta\theta_k \right).
\end{eqnarray}
In the long-range case with $\alpha<1$, though, the velocity \eqref{eq:gr_vel_max} diverges as $k^{\alpha-2}$ when $k\rightarrow 0$ and the short time behaviour of the entropy could be strongly influenced by the modes close to zero momenta if these are affected by the quench. However, we find for the quenches we plot in Fig.~\ref{fig:svnFermions1}, the change in the Bogoliubov angle $\delta\theta_k\rightarrow 0$, so the quench does not generate new quasiparticles at $k=0$, and existing quasiparticles just pick up a phase. Consequently, these play no role in the change of entanglement. On a formal level, the behaviour of $G$ (which ensures that the modes at $k=0$ are barely affected by the quench) is sufficient to ``cure'' the divergence of the velocity and the integrand in the right hand side of Eq.\eqref{eq:S_bound} becomes finite.

In Figs.~\ref{fig:svnFermions2}(a,b) a weak dependence on $\alpha$ for the qualitative behaviour of the entanglement entropy is shown clearly for this quench. In Fig.~\ref{fig:svnFermions2}(a) this is shown for $\mu=M/2$, and in Fig.~\ref{fig:svnFermions2}(b), this is shown for the subsystem in the bulk, $\mu\ll M$. 

However, if we choose a different initial state, for which the quasiparticles (as represented by the Bogoliubov angle) change substantially in the quench for $k\rightarrow 0$, then the behaviour can depend substantially on $\alpha$. In order to show this dependence --- analogous to what we saw in the previous section for the LRTI model, we need to consider ground states with a different structure than the $\Delta=4J$ ground state. Specifically, we look for situations where $\left| \varepsilon'(k) \right| G \left( \cos\delta\theta_k \right)$ in \eqref{eq:S} diverges as $k\rightarrow 0$, e.g., by choosing the ground state of a critical model as the starting point.  If we choose the ground state of the critical transverse-field Ising chain (with Hamiltonian $H_{\rm CTI}=\sum_l c^\dag_l (c^\dag_{l+1}+c_{l+1}-c_l)+\mathrm{h.c.}$), this has $\cos\theta_k^0\sim |k|/2$ and hence $G \left( \cos\delta\theta_k \right)=\log 2$. The behaviour of quasiparticles with momentum close to zero then dominates the dynamics at short times. 

Figs.~\ref{fig:svnFermions2}(c,d) show the entropy per unit length for the time evolution under the long-range Hamiltonian \eqref{eq:H_LRFH} with $\Delta=J/5$, again showing the half-chain entropy and a system in the bulk respectively. These calculations start from the ground state of the CTI Hamiltonian (with open boundary conditions). For $\alpha<1$ the short time behaviour is not dominated by the linear term and a completely different behaviour emerges. If we take the leading order terms at short times, we find that this behaviour is polynomial, and see that $S_\mu \propto t^{1/(2-\alpha)}$ at short times. The timescales over which this power law holds become shorter, and go to zero as $\alpha \rightarrow 1$. That is, in this case we have growth that is faster at very short times, but slower at intermediate times than the linear behaviour of short-range interactions. We see the same qualitative effects in the LRTI model treated in a Holstein-Primakoff approximation, as is detailed in Sec.~\ref{sec:HPapproximation}.

\section{Summary and Outlook}
\label{sec:summary}

We have compared the behaviour of correlation spreading and entanglement growth after global parameter quenches in the long-range transverse Ising model and a long-range fermion hopping model. In both models we find a clear delineation of long range behaviour for $\alpha<1$, and an intermediate regime where a light cone for correlation propagation becomes nonlinear and less well defined at $\alpha=2$. Counterintuitively, we find that for particular classes of initial states, growth of bipartite entanglement can be suppressed at short times for long-range interactions, because the structure of the initial state and the Hamiltonian symmetry prevent access to the full Hilbert space. We can characterise this change in entanglement growth based on the contribution of quasiparticles near $k=0$, where the group velocity and dispersion relation diverge for long-range interactions. At short times, we similarly see a suppression of correlation growth at long distances with long-range interactions. The results for the long-range transverse Ising model correspond directly to dynamics ongoing experiments with trapped ions. 

Near the completion of this work, we became aware of related studies, by Vodola et al. \cite{vodola2015}, of the ground state properties of the long-range fermionic hopping model, and by Van Regemortel et al.~\cite{Regemortel2015} of some dynamical properties related to those we studied here, in a model with a long-range pairing term.  

\begin{acknowledgments}
We thank A. Gorshkov, M. Kastner, and M. Van den Worm for stimulating discussions. This work was supported in part by AFOSR MURI FA9550-14-1-0035, and code development was supported by AFOSR grant FA9550-12-1-0057. The development of these ideas began at the Aspen Center for Physics, supported under NSF grant 1066293. Numerical calculations here utilised the ARCHIE-WeSt High Performance Computer.
\end{acknowledgments}

\bibliography{longrange}

\begin{thebibliography}{61}
\expandafter\ifx\csname natexlab\endcsname\relax\def\natexlab#1{#1}\fi
\expandafter\ifx\csname bibnamefont\endcsname\relax
  \def\bibnamefont#1{#1}\fi
\expandafter\ifx\csname bibfnamefont\endcsname\relax
  \def\bibfnamefont#1{#1}\fi
\expandafter\ifx\csname citenamefont\endcsname\relax
  \def\citenamefont#1{#1}\fi
\expandafter\ifx\csname url\endcsname\relax
  \def\url#1{\texttt{#1}}\fi
\expandafter\ifx\csname urlprefix\endcsname\relax\def\urlprefix{URL }\fi
\providecommand{\bibinfo}[2]{#2}
\providecommand{\eprint}[2][]{\url{#2}}

\bibitem[{\citenamefont{Schneider et~al.}(2012)\citenamefont{Schneider, Porras,
  and Schaetz}}]{Schneider2012}
\bibinfo{author}{\bibfnamefont{C.}~\bibnamefont{Schneider}},
  \bibinfo{author}{\bibfnamefont{D.}~\bibnamefont{Porras}}, \bibnamefont{and}
  \bibinfo{author}{\bibfnamefont{T.}~\bibnamefont{Schaetz}},
  \bibinfo{journal}{Reports on Progress in Physics}
  \textbf{\bibinfo{volume}{75}}, \bibinfo{pages}{024401}
  (\bibinfo{year}{2012}).

\bibitem[{\citenamefont{Islam et~al.}(2011)\citenamefont{Islam, Edwards, Kim,
  Korenblit, Noh, Carmichael, Lin, Duan, Joseph~Wang, Freericks
  et~al.}}]{Islam2011}
\bibinfo{author}{\bibfnamefont{R.}~\bibnamefont{Islam}},
  \bibinfo{author}{\bibfnamefont{E.~E.} \bibnamefont{Edwards}},
  \bibinfo{author}{\bibfnamefont{K.}~\bibnamefont{Kim}},
  \bibinfo{author}{\bibfnamefont{S.}~\bibnamefont{Korenblit}},
  \bibinfo{author}{\bibfnamefont{C.}~\bibnamefont{Noh}},
  \bibinfo{author}{\bibfnamefont{H.}~\bibnamefont{Carmichael}},
  \bibinfo{author}{\bibfnamefont{G.-D.} \bibnamefont{Lin}},
  \bibinfo{author}{\bibfnamefont{L.-M.} \bibnamefont{Duan}},
  \bibinfo{author}{\bibfnamefont{C.-C.} \bibnamefont{Joseph~Wang}},
  \bibinfo{author}{\bibfnamefont{J.~K.} \bibnamefont{Freericks}},
  \bibnamefont{et~al.}, \bibinfo{journal}{Nature Communications}
  \textbf{\bibinfo{volume}{2}}, \bibinfo{pages}{377} (\bibinfo{year}{2011}).

\bibitem[{\citenamefont{Kim et~al.}(2011)\citenamefont{Kim, Korenblit, Islam,
  Edwards, Chang, Noh, Carmichael, Lin, Duan, Wang et~al.}}]{Kim2011}
\bibinfo{author}{\bibfnamefont{K.}~\bibnamefont{Kim}},
  \bibinfo{author}{\bibfnamefont{S.}~\bibnamefont{Korenblit}},
  \bibinfo{author}{\bibfnamefont{R.}~\bibnamefont{Islam}},
  \bibinfo{author}{\bibfnamefont{E.~E.} \bibnamefont{Edwards}},
  \bibinfo{author}{\bibfnamefont{M.-S.} \bibnamefont{Chang}},
  \bibinfo{author}{\bibfnamefont{C.}~\bibnamefont{Noh}},
  \bibinfo{author}{\bibfnamefont{H.}~\bibnamefont{Carmichael}},
  \bibinfo{author}{\bibfnamefont{G.-D.} \bibnamefont{Lin}},
  \bibinfo{author}{\bibfnamefont{L.-M.} \bibnamefont{Duan}},
  \bibinfo{author}{\bibfnamefont{C.~C.~J.} \bibnamefont{Wang}},
  \bibnamefont{et~al.}, \bibinfo{journal}{New Journal of Physics}
  \textbf{\bibinfo{volume}{13}}, \bibinfo{pages}{105003}
  (\bibinfo{year}{2011}).

\bibitem[{\citenamefont{Lanyon et~al.}(2011)\citenamefont{Lanyon, Hempel, Nigg,
  M\"uller, Gerritsma, Z\"ahringer, Schindler, Barreiro, Rambach, Kirchmair
  et~al.}}]{Lanyon2011}
\bibinfo{author}{\bibfnamefont{B.~P.} \bibnamefont{Lanyon}},
  \bibinfo{author}{\bibfnamefont{C.}~\bibnamefont{Hempel}},
  \bibinfo{author}{\bibfnamefont{D.}~\bibnamefont{Nigg}},
  \bibinfo{author}{\bibfnamefont{M.}~\bibnamefont{M\"uller}},
  \bibinfo{author}{\bibfnamefont{R.}~\bibnamefont{Gerritsma}},
  \bibinfo{author}{\bibfnamefont{F.}~\bibnamefont{Z\"ahringer}},
  \bibinfo{author}{\bibfnamefont{P.}~\bibnamefont{Schindler}},
  \bibinfo{author}{\bibfnamefont{J.~T.} \bibnamefont{Barreiro}},
  \bibinfo{author}{\bibfnamefont{M.}~\bibnamefont{Rambach}},
  \bibinfo{author}{\bibfnamefont{G.}~\bibnamefont{Kirchmair}},
  \bibnamefont{et~al.}, \bibinfo{journal}{Science}
  \textbf{\bibinfo{volume}{334}}, \bibinfo{pages}{57} (\bibinfo{year}{2011}).

\bibitem[{\citenamefont{Blatt and Roos}(2012)}]{Blatt2012}
\bibinfo{author}{\bibfnamefont{R.}~\bibnamefont{Blatt}} \bibnamefont{and}
  \bibinfo{author}{\bibfnamefont{C.~F.} \bibnamefont{Roos}},
  \bibinfo{journal}{Nature Physics} \textbf{\bibinfo{volume}{8}},
  \bibinfo{pages}{277} (\bibinfo{year}{2012}).

\bibitem[{\citenamefont{Yan et~al.}(2013)\citenamefont{Yan, Moses, Gadway,
  Covey, Hazzard, Rey, Jin, and Ye}}]{Yan2013}
\bibinfo{author}{\bibfnamefont{B.}~\bibnamefont{Yan}},
  \bibinfo{author}{\bibfnamefont{S.~A.} \bibnamefont{Moses}},
  \bibinfo{author}{\bibfnamefont{B.}~\bibnamefont{Gadway}},
  \bibinfo{author}{\bibfnamefont{J.~P.} \bibnamefont{Covey}},
  \bibinfo{author}{\bibfnamefont{K.~R.~A.} \bibnamefont{Hazzard}},
  \bibinfo{author}{\bibfnamefont{A.~M.} \bibnamefont{Rey}},
  \bibinfo{author}{\bibfnamefont{D.~S.} \bibnamefont{Jin}}, \bibnamefont{and}
  \bibinfo{author}{\bibfnamefont{J.}~\bibnamefont{Ye}},
  \bibinfo{journal}{Nature} \textbf{\bibinfo{volume}{501}},
  \bibinfo{pages}{521} (\bibinfo{year}{2013}).

\bibitem[{\citenamefont{Micheli et~al.}(2006)\citenamefont{Micheli, Brennen,
  and Zoller}}]{Micheli2006}
\bibinfo{author}{\bibfnamefont{A.}~\bibnamefont{Micheli}},
  \bibinfo{author}{\bibfnamefont{G.~K.} \bibnamefont{Brennen}},
  \bibnamefont{and} \bibinfo{author}{\bibfnamefont{P.}~\bibnamefont{Zoller}},
  \bibinfo{journal}{Nat Phys} \textbf{\bibinfo{volume}{2}},
  \bibinfo{pages}{341} (\bibinfo{year}{2006}).

\bibitem[{\citenamefont{Schausz et~al.}(2012)\citenamefont{Schausz, Cheneau,
  Endres, Fukuhara, Hild, Omran, Pohl, Gross, Kuhr, and Bloch}}]{Schausz2012}
\bibinfo{author}{\bibfnamefont{P.}~\bibnamefont{Schausz}},
  \bibinfo{author}{\bibfnamefont{M.}~\bibnamefont{Cheneau}},
  \bibinfo{author}{\bibfnamefont{M.}~\bibnamefont{Endres}},
  \bibinfo{author}{\bibfnamefont{T.}~\bibnamefont{Fukuhara}},
  \bibinfo{author}{\bibfnamefont{S.}~\bibnamefont{Hild}},
  \bibinfo{author}{\bibfnamefont{A.}~\bibnamefont{Omran}},
  \bibinfo{author}{\bibfnamefont{T.}~\bibnamefont{Pohl}},
  \bibinfo{author}{\bibfnamefont{C.}~\bibnamefont{Gross}},
  \bibinfo{author}{\bibfnamefont{S.}~\bibnamefont{Kuhr}}, \bibnamefont{and}
  \bibinfo{author}{\bibfnamefont{I.}~\bibnamefont{Bloch}},
  \bibinfo{journal}{Nature} \textbf{\bibinfo{volume}{491}}, \bibinfo{pages}{87}
  (\bibinfo{year}{2012}).

\bibitem[{\citenamefont{Weimer et~al.}(2010)\citenamefont{Weimer, Muller,
  Lesanovsky, Zoller, and Buchler}}]{Weimer2010}
\bibinfo{author}{\bibfnamefont{H.}~\bibnamefont{Weimer}},
  \bibinfo{author}{\bibfnamefont{M.}~\bibnamefont{Muller}},
  \bibinfo{author}{\bibfnamefont{I.}~\bibnamefont{Lesanovsky}},
  \bibinfo{author}{\bibfnamefont{P.}~\bibnamefont{Zoller}}, \bibnamefont{and}
  \bibinfo{author}{\bibfnamefont{H.~P.} \bibnamefont{Buchler}},
  \bibinfo{journal}{Nat Phys} \textbf{\bibinfo{volume}{6}},
  \bibinfo{pages}{382} (\bibinfo{year}{2010}).

\bibitem[{\citenamefont{Pohl et~al.}(2010)\citenamefont{Pohl, Demler, and
  Lukin}}]{Pohl2010}
\bibinfo{author}{\bibfnamefont{T.}~\bibnamefont{Pohl}},
  \bibinfo{author}{\bibfnamefont{E.}~\bibnamefont{Demler}}, \bibnamefont{and}
  \bibinfo{author}{\bibfnamefont{M.~D.} \bibnamefont{Lukin}},
  \bibinfo{journal}{Phys. Rev. Lett.} \textbf{\bibinfo{volume}{104}},
  \bibinfo{pages}{043002} (\bibinfo{year}{2010}).

\bibitem[{\citenamefont{Porras and Cirac}(2004)}]{Porras2004}
\bibinfo{author}{\bibfnamefont{D.}~\bibnamefont{Porras}} \bibnamefont{and}
  \bibinfo{author}{\bibfnamefont{J.~I.} \bibnamefont{Cirac}},
  \bibinfo{journal}{Phys. Rev. Lett.} \textbf{\bibinfo{volume}{92}},
  \bibinfo{pages}{207901} (\bibinfo{year}{2004}).

\bibitem[{\citenamefont{Deng et~al.}(2005)\citenamefont{Deng, Porras, and
  Cirac}}]{Deng2005}
\bibinfo{author}{\bibfnamefont{X.-L.} \bibnamefont{Deng}},
  \bibinfo{author}{\bibfnamefont{D.}~\bibnamefont{Porras}}, \bibnamefont{and}
  \bibinfo{author}{\bibfnamefont{J.~I.} \bibnamefont{Cirac}},
  \bibinfo{journal}{Physical Review A} \textbf{\bibinfo{volume}{72}},
  \bibinfo{pages}{063407} (\bibinfo{year}{2005}).

\bibitem[{\citenamefont{Jurcevic et~al.}(2014)\citenamefont{Jurcevic, Lanyon,
  Hauke, Hempel, Zoller, Blatt, and Roos}}]{Jurcevic2014}
\bibinfo{author}{\bibfnamefont{P.}~\bibnamefont{Jurcevic}},
  \bibinfo{author}{\bibfnamefont{B.~P.} \bibnamefont{Lanyon}},
  \bibinfo{author}{\bibfnamefont{P.}~\bibnamefont{Hauke}},
  \bibinfo{author}{\bibfnamefont{C.}~\bibnamefont{Hempel}},
  \bibinfo{author}{\bibfnamefont{P.}~\bibnamefont{Zoller}},
  \bibinfo{author}{\bibfnamefont{R.}~\bibnamefont{Blatt}}, \bibnamefont{and}
  \bibinfo{author}{\bibfnamefont{C.~F.} \bibnamefont{Roos}},
  \bibinfo{journal}{Nature} \textbf{\bibinfo{volume}{511}},
  \bibinfo{pages}{202} (\bibinfo{year}{2014}).

\bibitem[{\citenamefont{Richerme et~al.}(2014)\citenamefont{Richerme, Gong,
  Lee, Senko, Smith, Foss-Feig, Michalakis, Gorshkov, and
  Monroe}}]{Richerme2014}
\bibinfo{author}{\bibfnamefont{P.}~\bibnamefont{Richerme}},
  \bibinfo{author}{\bibfnamefont{Z.-X.} \bibnamefont{Gong}},
  \bibinfo{author}{\bibfnamefont{A.}~\bibnamefont{Lee}},
  \bibinfo{author}{\bibfnamefont{C.}~\bibnamefont{Senko}},
  \bibinfo{author}{\bibfnamefont{J.}~\bibnamefont{Smith}},
  \bibinfo{author}{\bibfnamefont{M.}~\bibnamefont{Foss-Feig}},
  \bibinfo{author}{\bibfnamefont{S.}~\bibnamefont{Michalakis}},
  \bibinfo{author}{\bibfnamefont{A.~V.} \bibnamefont{Gorshkov}},
  \bibnamefont{and} \bibinfo{author}{\bibfnamefont{C.}~\bibnamefont{Monroe}},
  \bibinfo{journal}{Nature} \textbf{\bibinfo{volume}{511}},
  \bibinfo{pages}{198} (\bibinfo{year}{2014}).

\bibitem[{\citenamefont{Schachenmayer et~al.}(2013)\citenamefont{Schachenmayer,
  Lanyon, Roos, and Daley}}]{Schachenmayer2013}
\bibinfo{author}{\bibfnamefont{J.}~\bibnamefont{Schachenmayer}},
  \bibinfo{author}{\bibfnamefont{B.~P.} \bibnamefont{Lanyon}},
  \bibinfo{author}{\bibfnamefont{C.~F.} \bibnamefont{Roos}}, \bibnamefont{and}
  \bibinfo{author}{\bibfnamefont{A.~J.} \bibnamefont{Daley}},
  \bibinfo{journal}{Phys. Rev. X} \textbf{\bibinfo{volume}{3}},
  \bibinfo{pages}{031015} (\bibinfo{year}{2013}).

\bibitem[{\citenamefont{Hauke and Tagliacozzo}(2013)}]{Hauke2013}
\bibinfo{author}{\bibfnamefont{P.}~\bibnamefont{Hauke}} \bibnamefont{and}
  \bibinfo{author}{\bibfnamefont{L.}~\bibnamefont{Tagliacozzo}},
  \bibinfo{journal}{Phys. Rev. Lett.} \textbf{\bibinfo{volume}{111}},
  \bibinfo{pages}{207202} (\bibinfo{year}{2013}).

\bibitem[{\citenamefont{Gong and Duan}(2013)}]{Gong2013}
\bibinfo{author}{\bibfnamefont{Z.-X.} \bibnamefont{Gong}} \bibnamefont{and}
  \bibinfo{author}{\bibfnamefont{L.-M.} \bibnamefont{Duan}},
  \bibinfo{journal}{arXiv:1305.0985}  (\bibinfo{year}{2013}).

\bibitem[{\citenamefont{Britton et~al.}(2012)\citenamefont{Britton, Sawyer,
  Keith, Wang, Freericks, Uys, Biercuk, and Bollinger}}]{Britton2012}
\bibinfo{author}{\bibfnamefont{J.~W.} \bibnamefont{Britton}},
  \bibinfo{author}{\bibfnamefont{B.~C.} \bibnamefont{Sawyer}},
  \bibinfo{author}{\bibfnamefont{A.~C.} \bibnamefont{Keith}},
  \bibinfo{author}{\bibfnamefont{C.-C.~J.} \bibnamefont{Wang}},
  \bibinfo{author}{\bibfnamefont{J.~K.} \bibnamefont{Freericks}},
  \bibinfo{author}{\bibfnamefont{H.}~\bibnamefont{Uys}},
  \bibinfo{author}{\bibfnamefont{M.~J.} \bibnamefont{Biercuk}},
  \bibnamefont{and} \bibinfo{author}{\bibfnamefont{J.~J.}
  \bibnamefont{Bollinger}}, \bibinfo{journal}{Nature}
  \textbf{\bibinfo{volume}{484}}, \bibinfo{pages}{489} (\bibinfo{year}{2012}).

\bibitem[{\citenamefont{Lieb and Robinson}(1972)}]{Lieb1972}
\bibinfo{author}{\bibfnamefont{E.~H.} \bibnamefont{Lieb}} \bibnamefont{and}
  \bibinfo{author}{\bibfnamefont{D.~W.} \bibnamefont{Robinson}},
  \bibinfo{journal}{Comm. Math. Phys.} \textbf{\bibinfo{volume}{28}},
  \bibinfo{pages}{251} (\bibinfo{year}{1972}).

\bibitem[{\citenamefont{Hastings and Koma}(2006)}]{Hastings2006}
\bibinfo{author}{\bibfnamefont{M.}~\bibnamefont{Hastings}} \bibnamefont{and}
  \bibinfo{author}{\bibfnamefont{T.}~\bibnamefont{Koma}},
  \bibinfo{journal}{Communications in Mathematical Physics}
  \textbf{\bibinfo{volume}{265}}, \bibinfo{pages}{781} (\bibinfo{year}{2006}).

\bibitem[{\citenamefont{Eisert et~al.}(2013)\citenamefont{Eisert, van~den Worm,
  Manmana, and Kastner}}]{Eisert2013}
\bibinfo{author}{\bibfnamefont{J.}~\bibnamefont{Eisert}},
  \bibinfo{author}{\bibfnamefont{M.}~\bibnamefont{van~den Worm}},
  \bibinfo{author}{\bibfnamefont{S.~R.} \bibnamefont{Manmana}},
  \bibnamefont{and} \bibinfo{author}{\bibfnamefont{M.}~\bibnamefont{Kastner}},
  \bibinfo{journal}{Phys. Rev. Lett.} \textbf{\bibinfo{volume}{111}},
  \bibinfo{pages}{260401} (\bibinfo{year}{2013}).

\bibitem[{\citenamefont{Gong et~al.}(2014)\citenamefont{Gong, Foss-Feig,
  Michalakis, and Gorshkov}}]{Gong2014}
\bibinfo{author}{\bibfnamefont{Z.-X.} \bibnamefont{Gong}},
  \bibinfo{author}{\bibfnamefont{M.}~\bibnamefont{Foss-Feig}},
  \bibinfo{author}{\bibfnamefont{S.}~\bibnamefont{Michalakis}},
  \bibnamefont{and} \bibinfo{author}{\bibfnamefont{A.~V.}
  \bibnamefont{Gorshkov}}, \bibinfo{journal}{Phys. Rev. Lett.}
  \textbf{\bibinfo{volume}{113}}, \bibinfo{pages}{030602}
  (\bibinfo{year}{2014}).

\bibitem[{\citenamefont{Storch et~al.}(2015)\citenamefont{Storch, van~den Worm,
  and Kastner}}]{Storch2015}
\bibinfo{author}{\bibfnamefont{D.-M.} \bibnamefont{Storch}},
  \bibinfo{author}{\bibfnamefont{M.}~\bibnamefont{van~den Worm}},
  \bibnamefont{and} \bibinfo{author}{\bibfnamefont{M.}~\bibnamefont{Kastner}},
  \bibinfo{journal}{New Journal of Physics} \textbf{\bibinfo{volume}{17}},
  \bibinfo{pages}{063021} (\bibinfo{year}{2015}).

\bibitem[{\citenamefont{Foss-Feig et~al.}(2015)\citenamefont{Foss-Feig, Gong,
  Clark, and Gorshkov}}]{Foss-Feig2015}
\bibinfo{author}{\bibfnamefont{M.}~\bibnamefont{Foss-Feig}},
  \bibinfo{author}{\bibfnamefont{Z.-X.} \bibnamefont{Gong}},
  \bibinfo{author}{\bibfnamefont{C.~W.} \bibnamefont{Clark}}, \bibnamefont{and}
  \bibinfo{author}{\bibfnamefont{A.~V.} \bibnamefont{Gorshkov}},
  \bibinfo{journal}{Phys. Rev. Lett.} \textbf{\bibinfo{volume}{114}},
  \bibinfo{pages}{157201} (\bibinfo{year}{2015}).

\bibitem[{\citenamefont{Maghrebi et~al.}(2015)\citenamefont{Maghrebi, Gong,
  Foss-Feig, and Gorshkov}}]{Maghrebi2015}
\bibinfo{author}{\bibfnamefont{M.~F.} \bibnamefont{Maghrebi}},
  \bibinfo{author}{\bibfnamefont{Z.-X.} \bibnamefont{Gong}},
  \bibinfo{author}{\bibfnamefont{M.}~\bibnamefont{Foss-Feig}},
  \bibnamefont{and} \bibinfo{author}{\bibfnamefont{A.~V.}
  \bibnamefont{Gorshkov}}, \bibinfo{journal}{arXiv:1508.00906}
  (\bibinfo{year}{2015}).

\bibitem[{\citenamefont{Schachenmayer et~al.}(2015)\citenamefont{Schachenmayer,
  Pikovski, and Rey}}]{Schachenmayer2015}
\bibinfo{author}{\bibfnamefont{J.}~\bibnamefont{Schachenmayer}},
  \bibinfo{author}{\bibfnamefont{A.}~\bibnamefont{Pikovski}}, \bibnamefont{and}
  \bibinfo{author}{\bibfnamefont{A.~M.} \bibnamefont{Rey}},
  \bibinfo{journal}{New Journal of Physics} \textbf{\bibinfo{volume}{17}},
  \bibinfo{pages}{065009} (\bibinfo{year}{2015}).

\bibitem[{\citenamefont{Rajabpour and Sotiriadis}(2014)}]{Rajabpour2014}
\bibinfo{author}{\bibfnamefont{M.~A.} \bibnamefont{Rajabpour}}
  \bibnamefont{and}
  \bibinfo{author}{\bibfnamefont{S.}~\bibnamefont{Sotiriadis}},
  \bibinfo{journal}{arXiv:1409.6558}  (\bibinfo{year}{2014}).

\bibitem[{\citenamefont{Hazzard et~al.}(2014)\citenamefont{Hazzard, van~den
  Worm, Foss-Feig, Manmana, Dalla~Torre, Pfau, Kastner, and Rey}}]{Hazzard2014}
\bibinfo{author}{\bibfnamefont{K.~R.~A.} \bibnamefont{Hazzard}},
  \bibinfo{author}{\bibfnamefont{M.}~\bibnamefont{van~den Worm}},
  \bibinfo{author}{\bibfnamefont{M.}~\bibnamefont{Foss-Feig}},
  \bibinfo{author}{\bibfnamefont{S.~R.} \bibnamefont{Manmana}},
  \bibinfo{author}{\bibfnamefont{E.~G.} \bibnamefont{Dalla~Torre}},
  \bibinfo{author}{\bibfnamefont{T.}~\bibnamefont{Pfau}},
  \bibinfo{author}{\bibfnamefont{M.}~\bibnamefont{Kastner}}, \bibnamefont{and}
  \bibinfo{author}{\bibfnamefont{A.~M.} \bibnamefont{Rey}},
  \bibinfo{journal}{Phys. Rev. A} \textbf{\bibinfo{volume}{90}},
  \bibinfo{pages}{063622} (\bibinfo{year}{2014}).

\bibitem[{\citenamefont{Cevolani et~al.}(2015)\citenamefont{Cevolani, Carleo,
  and Sanchez-Palencia}}]{Cevolani2015}
\bibinfo{author}{\bibfnamefont{L.}~\bibnamefont{Cevolani}},
  \bibinfo{author}{\bibfnamefont{G.}~\bibnamefont{Carleo}}, \bibnamefont{and}
  \bibinfo{author}{\bibfnamefont{L.}~\bibnamefont{Sanchez-Palencia}},
  \bibinfo{journal}{arXiv:1503.01786}  (\bibinfo{year}{2015}).

\bibitem[{\citenamefont{Tezuka et~al.}(2014)\citenamefont{Tezuka,
  Garc\'{\i}a-Garc\'{\i}a, and Cazalilla}}]{Tezuka2014}
\bibinfo{author}{\bibfnamefont{M.}~\bibnamefont{Tezuka}},
  \bibinfo{author}{\bibfnamefont{A.~M.} \bibnamefont{Garc\'{\i}a-Garc\'{\i}a}},
  \bibnamefont{and} \bibinfo{author}{\bibfnamefont{M.~A.}
  \bibnamefont{Cazalilla}}, \bibinfo{journal}{Phys. Rev. A}
  \textbf{\bibinfo{volume}{90}}, \bibinfo{pages}{053618}
  (\bibinfo{year}{2014}).

\bibitem[{\citenamefont{Kitaev}(2001)}]{Kitaev2001}
\bibinfo{author}{\bibfnamefont{A.~Y.} \bibnamefont{Kitaev}},
  \bibinfo{journal}{Physics-Uspekhi} \textbf{\bibinfo{volume}{44}},
  \bibinfo{pages}{131} (\bibinfo{year}{2001}).

\bibitem[{\citenamefont{Verstraete et~al.}(2008)\citenamefont{Verstraete, Murg,
  and Cirac}}]{Verstraete2008}
\bibinfo{author}{\bibfnamefont{F.}~\bibnamefont{Verstraete}},
  \bibinfo{author}{\bibfnamefont{V.}~\bibnamefont{Murg}}, \bibnamefont{and}
  \bibinfo{author}{\bibfnamefont{J.~I.} \bibnamefont{Cirac}},
  \bibinfo{journal}{Advances in Physics} \textbf{\bibinfo{volume}{57}},
  \bibinfo{pages}{143 } (\bibinfo{year}{2008}).

\bibitem[{\citenamefont{Garc{\'\i}a-Ripoll}(2006)}]{Garcia-Ripoll2006}
\bibinfo{author}{\bibfnamefont{J.~J.} \bibnamefont{Garc{\'\i}a-Ripoll}},
  \bibinfo{journal}{New Journal of Physics} \textbf{\bibinfo{volume}{8}},
  \bibinfo{pages}{305} (\bibinfo{year}{2006}).

\bibitem[{\citenamefont{McCulloch}(2007)}]{McCulloch2007}
\bibinfo{author}{\bibfnamefont{I.~P.} \bibnamefont{McCulloch}},
  \bibinfo{journal}{Journal of Statistical Mechanics: Theory and Experiment}
  \textbf{\bibinfo{volume}{2007}}, \bibinfo{pages}{P10014}
  (\bibinfo{year}{2007}).

\bibitem[{\citenamefont{Pirvu et~al.}(2010)\citenamefont{Pirvu, Murg, Cirac,
  and Verstraete}}]{Pirvu2010}
\bibinfo{author}{\bibfnamefont{B.}~\bibnamefont{Pirvu}},
  \bibinfo{author}{\bibfnamefont{V.}~\bibnamefont{Murg}},
  \bibinfo{author}{\bibfnamefont{J.~I.} \bibnamefont{Cirac}}, \bibnamefont{and}
  \bibinfo{author}{\bibfnamefont{F.}~\bibnamefont{Verstraete}},
  \bibinfo{journal}{New Journal of Physics} \textbf{\bibinfo{volume}{12}},
  \bibinfo{pages}{025012} (\bibinfo{year}{2010}).

\bibitem[{\citenamefont{Crosswhite et~al.}(2008)\citenamefont{Crosswhite,
  Doherty, and Vidal}}]{Crosswhite2008}
\bibinfo{author}{\bibfnamefont{G.~M.} \bibnamefont{Crosswhite}},
  \bibinfo{author}{\bibfnamefont{A.~C.} \bibnamefont{Doherty}},
  \bibnamefont{and} \bibinfo{author}{\bibfnamefont{G.}~\bibnamefont{Vidal}},
  \bibinfo{journal}{Physical Review B} \textbf{\bibinfo{volume}{78}},
  \bibinfo{pages}{035116} (\bibinfo{year}{2008}).

\bibitem[{\citenamefont{Schollwoeck}(2011)}]{Schollwoeck2011}
\bibinfo{author}{\bibfnamefont{U.}~\bibnamefont{Schollwoeck}},
  \bibinfo{journal}{Annals of Physics} \textbf{\bibinfo{volume}{326}},
  \bibinfo{pages}{96} (\bibinfo{year}{2011}).

\bibitem[{\citenamefont{Schollwoeck}(2005)}]{Schollwoeck2005}
\bibinfo{author}{\bibfnamefont{U.}~\bibnamefont{Schollwoeck}},
  \bibinfo{journal}{Rev. Mod. Phys.} \textbf{\bibinfo{volume}{77}},
  \bibinfo{pages}{259} (\bibinfo{year}{2005}).

\bibitem[{\citenamefont{White and Feiguin}(2004)}]{White2004}
\bibinfo{author}{\bibfnamefont{S.~R.} \bibnamefont{White}} \bibnamefont{and}
  \bibinfo{author}{\bibfnamefont{A.~E.} \bibnamefont{Feiguin}},
  \bibinfo{journal}{Phys. Rev. Lett.} \textbf{\bibinfo{volume}{93}},
  \bibinfo{pages}{076401} (\bibinfo{year}{2004}).

\bibitem[{\citenamefont{Daley et~al.}(2004)\citenamefont{Daley, Kollath,
  Schollwock, and Vidal}}]{Daley2004}
\bibinfo{author}{\bibfnamefont{A.~J.} \bibnamefont{Daley}},
  \bibinfo{author}{\bibfnamefont{C.}~\bibnamefont{Kollath}},
  \bibinfo{author}{\bibfnamefont{U.}~\bibnamefont{Schollwock}},
  \bibnamefont{and} \bibinfo{author}{\bibfnamefont{G.}~\bibnamefont{Vidal}},
  \bibinfo{journal}{Journal of Statistical Mechanics: Theory and Experiment}
  \textbf{\bibinfo{volume}{2004}}, \bibinfo{pages}{P04005}
  (\bibinfo{year}{2004}).

\bibitem[{\citenamefont{Haegeman et~al.}(2014)\citenamefont{Haegeman, Lubich,
  Oseledets, Vandereycken, and Verstraete}}]{Haegeman2014}
\bibinfo{author}{\bibfnamefont{J.}~\bibnamefont{Haegeman}},
  \bibinfo{author}{\bibfnamefont{C.}~\bibnamefont{Lubich}},
  \bibinfo{author}{\bibfnamefont{I.}~\bibnamefont{Oseledets}},
  \bibinfo{author}{\bibfnamefont{B.}~\bibnamefont{Vandereycken}},
  \bibnamefont{and}
  \bibinfo{author}{\bibfnamefont{F.}~\bibnamefont{Verstraete}},
  \bibinfo{journal}{arXiv:1408.5056}  (\bibinfo{year}{2014}).

\bibitem[{\citenamefont{Vodola et~al.}(2015)\citenamefont{Vodola, Lepori,
  Ercolessi, and Pupillo}}]{vodola2015}
\bibinfo{author}{\bibfnamefont{D.}~\bibnamefont{Vodola}},
  \bibinfo{author}{\bibfnamefont{L.}~\bibnamefont{Lepori}},
  \bibinfo{author}{\bibfnamefont{E.}~\bibnamefont{Ercolessi}},
  \bibnamefont{and} \bibinfo{author}{\bibfnamefont{G.}~\bibnamefont{Pupillo}},
  \bibinfo{journal}{arXiv:1508.00820}  (\bibinfo{year}{2015}).

\bibitem[{\citenamefont{Cheneau et~al.}(2012)\citenamefont{Cheneau, Barmettler,
  Poletti, Endres, Schauß, Fukuhara, Gross, Bloch, Kollath, and
  Kuhr}}]{Cheneau2012}
\bibinfo{author}{\bibfnamefont{M.}~\bibnamefont{Cheneau}},
  \bibinfo{author}{\bibfnamefont{P.}~\bibnamefont{Barmettler}},
  \bibinfo{author}{\bibfnamefont{D.}~\bibnamefont{Poletti}},
  \bibinfo{author}{\bibfnamefont{M.}~\bibnamefont{Endres}},
  \bibinfo{author}{\bibfnamefont{P.}~\bibnamefont{Schauß}},
  \bibinfo{author}{\bibfnamefont{T.}~\bibnamefont{Fukuhara}},
  \bibinfo{author}{\bibfnamefont{C.}~\bibnamefont{Gross}},
  \bibinfo{author}{\bibfnamefont{I.}~\bibnamefont{Bloch}},
  \bibinfo{author}{\bibfnamefont{C.}~\bibnamefont{Kollath}}, \bibnamefont{and}
  \bibinfo{author}{\bibfnamefont{S.}~\bibnamefont{Kuhr}},
  \bibinfo{journal}{Nature} \textbf{\bibinfo{volume}{481}},
  \bibinfo{pages}{484} (\bibinfo{year}{2012}).

\bibitem[{\citenamefont{Calabrese et~al.}(2011)\citenamefont{Calabrese, Essler,
  and Fagotti}}]{Calabrese2011}
\bibinfo{author}{\bibfnamefont{P.}~\bibnamefont{Calabrese}},
  \bibinfo{author}{\bibfnamefont{F.~H.~L.} \bibnamefont{Essler}},
  \bibnamefont{and} \bibinfo{author}{\bibfnamefont{M.}~\bibnamefont{Fagotti}},
  \bibinfo{journal}{Phys. Rev. Lett.} \textbf{\bibinfo{volume}{106}},
  \bibinfo{pages}{227203} (\bibinfo{year}{2011}).

\bibitem[{\citenamefont{Calabrese and Cardy}(2005)}]{Calabrese2005}
\bibinfo{author}{\bibfnamefont{P.}~\bibnamefont{Calabrese}} \bibnamefont{and}
  \bibinfo{author}{\bibfnamefont{J.}~\bibnamefont{Cardy}},
  \bibinfo{journal}{Journal of Statistical Mechanics: Theory and Experiment}
  \textbf{\bibinfo{volume}{2005}}, \bibinfo{pages}{P04010}
  (\bibinfo{year}{2005}).

\bibitem[{\citenamefont{Fagotti and Calabrese}(2008)}]{Fagotti2008}
\bibinfo{author}{\bibfnamefont{M.}~\bibnamefont{Fagotti}} \bibnamefont{and}
  \bibinfo{author}{\bibfnamefont{P.}~\bibnamefont{Calabrese}},
  \bibinfo{journal}{Phys. Rev. A} \textbf{\bibinfo{volume}{78}},
  \bibinfo{pages}{010306} (\bibinfo{year}{2008}).

\bibitem[{\citenamefont{Igl\'oi and Rieger}(2000)}]{Igloi2000}
\bibinfo{author}{\bibfnamefont{F.}~\bibnamefont{Igl\'oi}} \bibnamefont{and}
  \bibinfo{author}{\bibfnamefont{H.}~\bibnamefont{Rieger}},
  \bibinfo{journal}{Phys. Rev. Lett.} \textbf{\bibinfo{volume}{85}},
  \bibinfo{pages}{3233} (\bibinfo{year}{2000}).

\bibitem[{\citenamefont{Rossini et~al.}(2009)\citenamefont{Rossini, Silva,
  Mussardo, and Santoro}}]{Rossini2009}
\bibinfo{author}{\bibfnamefont{D.}~\bibnamefont{Rossini}},
  \bibinfo{author}{\bibfnamefont{A.}~\bibnamefont{Silva}},
  \bibinfo{author}{\bibfnamefont{G.}~\bibnamefont{Mussardo}}, \bibnamefont{and}
  \bibinfo{author}{\bibfnamefont{G.~E.} \bibnamefont{Santoro}},
  \bibinfo{journal}{Phys. Rev. Lett.} \textbf{\bibinfo{volume}{102}},
  \bibinfo{pages}{127204} (\bibinfo{year}{2009}).

\bibitem[{\citenamefont{Rossini et~al.}(2010)\citenamefont{Rossini, Suzuki,
  Mussardo, Santoro, and Silva}}]{Rossini2010}
\bibinfo{author}{\bibfnamefont{D.}~\bibnamefont{Rossini}},
  \bibinfo{author}{\bibfnamefont{S.}~\bibnamefont{Suzuki}},
  \bibinfo{author}{\bibfnamefont{G.}~\bibnamefont{Mussardo}},
  \bibinfo{author}{\bibfnamefont{G.~E.} \bibnamefont{Santoro}},
  \bibnamefont{and} \bibinfo{author}{\bibfnamefont{A.}~\bibnamefont{Silva}},
  \bibinfo{journal}{Phys. Rev. B} \textbf{\bibinfo{volume}{82}},
  \bibinfo{pages}{144302} (\bibinfo{year}{2010}).

\bibitem[{\citenamefont{Rieger and Igl\'oi}(2011)}]{Rieger2011}
\bibinfo{author}{\bibfnamefont{H.}~\bibnamefont{Rieger}} \bibnamefont{and}
  \bibinfo{author}{\bibfnamefont{F.}~\bibnamefont{Igl\'oi}},
  \bibinfo{journal}{Phys. Rev. B} \textbf{\bibinfo{volume}{84}},
  \bibinfo{pages}{165117} (\bibinfo{year}{2011}).

\bibitem[{\citenamefont{Calabrese
  et~al.}(2012{\natexlab{a}})\citenamefont{Calabrese, Essler, and
  Fagotti}}]{Calabrese2012a}
\bibinfo{author}{\bibfnamefont{P.}~\bibnamefont{Calabrese}},
  \bibinfo{author}{\bibfnamefont{F.~H.~L.} \bibnamefont{Essler}},
  \bibnamefont{and} \bibinfo{author}{\bibfnamefont{M.}~\bibnamefont{Fagotti}},
  \bibinfo{journal}{Journal of Statistical Mechanics: Theory and Experiment}
  \textbf{\bibinfo{volume}{2012}}, \bibinfo{pages}{P07016}
  (\bibinfo{year}{2012}{\natexlab{a}}).

\bibitem[{\citenamefont{Calabrese
  et~al.}(2012{\natexlab{b}})\citenamefont{Calabrese, Essler, and
  Fagotti}}]{Calabrese2012b}
\bibinfo{author}{\bibfnamefont{P.}~\bibnamefont{Calabrese}},
  \bibinfo{author}{\bibfnamefont{F.~H.~L.} \bibnamefont{Essler}},
  \bibnamefont{and} \bibinfo{author}{\bibfnamefont{M.}~\bibnamefont{Fagotti}},
  \bibinfo{journal}{Journal of Statistical Mechanics: Theory and Experiment}
  \textbf{\bibinfo{volume}{2012}}, \bibinfo{pages}{P07022}
  (\bibinfo{year}{2012}{\natexlab{b}}).

\bibitem[{\citenamefont{Schuricht and Essler}(2012)}]{Schuricht2012}
\bibinfo{author}{\bibfnamefont{D.}~\bibnamefont{Schuricht}} \bibnamefont{and}
  \bibinfo{author}{\bibfnamefont{F.~H.~L.} \bibnamefont{Essler}},
  \bibinfo{journal}{Journal of Statistical Mechanics: Theory and Experiment}
  \textbf{\bibinfo{volume}{2012}}, \bibinfo{pages}{P04017}
  (\bibinfo{year}{2012}).

\bibitem[{\citenamefont{Foini et~al.}(2012)\citenamefont{Foini, Cugliandolo,
  and Gambassi}}]{Foini2012}
\bibinfo{author}{\bibfnamefont{L.}~\bibnamefont{Foini}},
  \bibinfo{author}{\bibfnamefont{L.~F.} \bibnamefont{Cugliandolo}},
  \bibnamefont{and} \bibinfo{author}{\bibfnamefont{A.}~\bibnamefont{Gambassi}},
  \bibinfo{journal}{Journal of Statistical Mechanics: Theory and Experiment}
  \textbf{\bibinfo{volume}{2012}}, \bibinfo{pages}{P09011}
  (\bibinfo{year}{2012}).

\bibitem[{\citenamefont{Essler et~al.}(2012)\citenamefont{Essler, Evangelisti,
  and Fagotti}}]{Essler2012}
\bibinfo{author}{\bibfnamefont{F.~H.~L.} \bibnamefont{Essler}},
  \bibinfo{author}{\bibfnamefont{S.}~\bibnamefont{Evangelisti}},
  \bibnamefont{and} \bibinfo{author}{\bibfnamefont{M.}~\bibnamefont{Fagotti}},
  \bibinfo{journal}{Phys. Rev. Lett.} \textbf{\bibinfo{volume}{109}},
  \bibinfo{pages}{247206} (\bibinfo{year}{2012}).

\bibitem[{\citenamefont{Fagotti and Essler}(2013)}]{Fagotti2013}
\bibinfo{author}{\bibfnamefont{M.}~\bibnamefont{Fagotti}} \bibnamefont{and}
  \bibinfo{author}{\bibfnamefont{F.~H.~L.} \bibnamefont{Essler}},
  \bibinfo{journal}{Phys. Rev. B} \textbf{\bibinfo{volume}{87}},
  \bibinfo{pages}{245107} (\bibinfo{year}{2013}).

\bibitem[{\citenamefont{Caux and Essler}(2013)}]{Caux2013}
\bibinfo{author}{\bibfnamefont{J.-S.} \bibnamefont{Caux}} \bibnamefont{and}
  \bibinfo{author}{\bibfnamefont{F.~H.~L.} \bibnamefont{Essler}},
  \bibinfo{journal}{Phys. Rev. Lett.} \textbf{\bibinfo{volume}{110}},
  \bibinfo{pages}{257203} (\bibinfo{year}{2013}).

\bibitem[{\citenamefont{Vidal et~al.}(2004)\citenamefont{Vidal, Palacios, and
  Aslangul}}]{Vidal2004}
\bibinfo{author}{\bibfnamefont{J.}~\bibnamefont{Vidal}},
  \bibinfo{author}{\bibfnamefont{G.}~\bibnamefont{Palacios}}, \bibnamefont{and}
  \bibinfo{author}{\bibfnamefont{C.}~\bibnamefont{Aslangul}},
  \bibinfo{journal}{Phys. Rev. A} \textbf{\bibinfo{volume}{70}},
  \bibinfo{pages}{062304} (\bibinfo{year}{2004}).

\bibitem[{\citenamefont{Wilms et~al.}(2012)\citenamefont{Wilms, Vidal,
  Verstraete, and Dusuel}}]{Wilms2012}
\bibinfo{author}{\bibfnamefont{J.}~\bibnamefont{Wilms}},
  \bibinfo{author}{\bibfnamefont{J.}~\bibnamefont{Vidal}},
  \bibinfo{author}{\bibfnamefont{F.}~\bibnamefont{Verstraete}},
  \bibnamefont{and} \bibinfo{author}{\bibfnamefont{S.}~\bibnamefont{Dusuel}},
  \bibinfo{journal}{Journal of Statistical Mechanics: Theory and Experiment}
  \textbf{\bibinfo{volume}{2012}}, \bibinfo{pages}{P01023}
  (\bibinfo{year}{2012}).

\bibitem[{\citenamefont{Latorre et~al.}(2005)\citenamefont{Latorre, Or\'us,
  Rico, and Vidal}}]{Latorre2005}
\bibinfo{author}{\bibfnamefont{J.~I.} \bibnamefont{Latorre}},
  \bibinfo{author}{\bibfnamefont{R.}~\bibnamefont{Or\'us}},
  \bibinfo{author}{\bibfnamefont{E.}~\bibnamefont{Rico}}, \bibnamefont{and}
  \bibinfo{author}{\bibfnamefont{J.}~\bibnamefont{Vidal}},
  \bibinfo{journal}{Phys. Rev. A} \textbf{\bibinfo{volume}{71}},
  \bibinfo{pages}{064101} (\bibinfo{year}{2005}).

\bibitem[{\citenamefont{Regemortel et~al.}(2015)\citenamefont{Regemortel, Sels,
  and Wouters}}]{Regemortel2015}
\bibinfo{author}{\bibfnamefont{M.~V.} \bibnamefont{Regemortel}},
  \bibinfo{author}{\bibfnamefont{D.}~\bibnamefont{Sels}}, \bibnamefont{and}
  \bibinfo{author}{\bibfnamefont{M.}~\bibnamefont{Wouters}},
  \bibinfo{journal}{arXiv:1511.05459}  (\bibinfo{year}{2015}).

\end{thebibliography}

\end{document}